\documentstyle[12pt,epsf]{article}
\def\del        {  \partial  }
\def\half       {  {1\over 2}  }

\def\defint#1#2 {  \int_{#1}^{#2}  }
\def\rootof#1   {  \left( #1 \right)^{1/2}  } 
\def\deldel#1   {  {\partial\over \partial #1}  }
\def\trace      {  \mbox{Tr}  }
\def\abs#1      {  \vert #1 \vert  }
\def\ie         {  {\it i.e.}      }
\def\evalat#1   {  \left\vert_{#1} \right. } 
\def\e          { {\rm e}  }

\def\period     { \, . }
\def\comma      { \, , }
\def\lsim       { \lower .65ex \hbox{\ $\stackrel{<}{\sim}$\ } }
\def\gsim       { \lower .65ex \hbox{\ $\stackrel{>}{\sim}$\ } }

\def\calO       { {\cal O} }

\def\calR       { {\cal R} }

\def\vecii#1#2      {  \left(\begin{array}{c}#1\\#2\end{array}\right)  }
\def\veciii#1#2#3   {  \left(\begin{array}{c}#1\\#2\\#3\end{array}\right)  }
\def\veciv#1#2#3#4  {  \left(\begin{array}{c}#1\\#2\\#3\\#4
                                 \end{array}\right)  }
\def\vecv#1#2#3#4#5 {  \left(\begin{array}{c}#1\\#2\\#3\\#4\\#5
                                 \end{array}\right)  }

\def\matrixii#1#2#3#4            {  \left(\begin{array}{cc}#1&#2\\#3&#4
                                       \end{array}\right) }
\def\matrixiii#1#2#3#4#5#6#7#8#9 {  \left(\begin{array}{ccc}#1&#2&#3\\
                                     #4&#5&#6\\#7&#8&#9\end{array}\right)  }
\def\mativ#1#2#3#4               {  \left(\begin{array}{cccc}
                                       #1\\#2\\#3\\#4\end{array}\right) }
\def\matv#1#2#3#4#5              {  \left(\begin{array}{ccccc}
                                     #1\\#2\\#3\\#4\\#5\end{array}\right)  }
\def\eqabegin         {  \begin{eqnarray}  }
\def\eqaend           {  \end{eqnarray}  }
\def\nn               {  \nonumber  }
\def\bracetwo#1#2     {  \left\{ \begin{array}{l} #1 \\ #2 \end{array}
                         \right.  }
\def\bracetwocases#1#2#3#4  {   \left\{ \begin{array}{ll} #1 & \qquad #2 \\
                                 #3 & \qquad #4 \end{array} \right.  }
\def\bracebegin#1     {  \left\{ \begin{array}{#1}   }
\def\braceend         {  \end{array}\right.   }
\def\parn                 { \par\noindent  }
\def\parbigskip        {  \par\bigskip  }
\def\parmedskip        {  \par\medskip  }
\def\parsmallskip      {  \par\smallskip  }
\def\parbigskipn      { \parbigskip\noindent }
\def\parmedskipn     { \parmedskip\noindent }
\def\parsmallskipn    { \parsmallskip\noindent }

\def\parag#1           {\paragraph{#1} \mbox{ }\parmedskip\noindent}
\def\boxit#1#2      {  \vbox{\hrule\hbox{ \hskip -4.1pt \vrule\kern3pt \vbox
                    {  \hsize #1 \strut\kern3pt #2 \kern3pt\strut  }
                       \kern3pt  \vrule} \hrule  } }
\def\centerbox#1#2  {  \mbox{  }\par\bigskip  \hfil \boxit{#1}{#2} \hfil
                       \par\bigskip\noindent }
\def\rightbox#1#2   {  \hfill\boxit{#1}{#2}  }
\def\leftbox#1#2    {  \boxit{#1}{#2}  }
\def\fullbox#1      {  \boxit{\textwidth}{#1}  }
\def\trianglemap#1#2#3#4#5#6  {   {\large $$ \begin{array}{rcl} #1\!\!\!
            &{\stackrel{{\scriptstyle #2}}{\longrightarrow   }}&\!\!\!  #3 \\ 
                                  { } & {\scriptstyle #4}\!\!\!\searrow \quad
                                  \swarrow \!\!\!{\scriptstyle #5}& { } \\
                                  { } & #6 & { } \end{array} $$ }    }
\def\squaremap#1#2#3#4#5#6#7#8    { {\large $$ \begin{array}{ccc}#1 &
                      \stackrel{{\scriptstyle #2}}{\longrightarrow} & #3 \\
                      {\scriptstyle #4}\!\downarrow & { } & \downarrow \!
                      {\scriptstyle #5}\\ #6 &\!\!
                      \longrightarrow_{{ }_{\!\!\!\!\!\!\!\!\!\!\!
                      {\scriptstyle #7}}}    &#8 \end{array} $$ }   }
\def\righttrianglemap#1#2#3#4#5#6  {  {\large $$ \begin{array}{rcl}
                      #1\!\! & \stackrel{{\scriptstyle #2}}{\longrightarrow} 
                      & #3 \\  { }&\!\!{\scriptstyle #4}\!\!\searrow
                      & \downarrow \!\!{\scriptstyle #5}\\
                      { }&{ }& #6 \end{array} $$ }   }
\def\figspace#1            {  \par\vspace{#1}\noindent  }
\def\topfigspace#1       {  \par\vspace*{#1}\noindent  }
\def\rightfigspacebegin  {  \par\noindent\begin{minipage}[t]{10cm}  }
\def\rightfigspaceend    {  \end{minipage}\par\noindent  }
\def\leftfigspacebegin   {  \par\noindent
                             \hspace*{10cm}\begin{minipage}[t]{6cm} }
\def\leftfigspaceend     {  \end{minipage}\par\noindent  }
\def\titleandfile#1#2   {  \begin{center}{\Large\bf #1}\end{center}
                            \par\begin{flushright} #2 \end{flushright}  }
\def\msection#1      {  \begin{center} \section{#1} \end{center}   }
\def\nsection#1      {  \let\boldface\bf \def\bf{} \section{#1}
                           \let\bf\boldface   }
\def\mnsection#1     {  \begin{center} \nsection{#1} \end{center}  }
\def\capsection#1    {  \let\boldface\bf \def\bf{\sc} \section{#1}
                           \let\bf\boldface   }
\def\mcapsection#1   {  \begin{center} \capsection{#1} \end{center} }
\def\sectionnumbering{\setcounter{equation}{0}
          \renewcommand{\theequation}{\arabic{section}.\arabic{equation}}}
\def\nullify#1 {}
\def\period{\, .}
\def\comma{\, ,}

\def\overrttwo{{1 \over \sqrt{2}}}
\def\overi{{1\over i}}
\def\ff{\gamma^2}  
\def\xplus{{x^+}}
\def\xminus{{x^-}}
\def\yplus{{y^+}}

\def\delplus{\del_+}
\def\delminus{\del_-}

\def\deltamn{\delta_{m+n,0}}
\def\ket#1{\mid #1 >}
\def\bra#1{< #1 \mid }

\def\rhodot{\dot{\rho}}

\def\Phidot{\dot{\Phi}}

\def\ghat{\hat{g}}

\def\deltamn{\delta_{m+n,0}}

\def\downvacket{\mid 0 >_\downarrow}
\def\pdownvac{\mid \vec{P} >_\downarrow}

\def\psmearvac{\mid \tilde{P} >} 
\def\psmearnorm{ < \tilde{P} \mid  \tilde{P} >} 
\def\Psizeroket{\mid \Psi_0 >}
\def\weight{W(p^+)}
  
        \def\Ldl{L^{dL}}
            \def\Lf{L^f}

\def\Pplus{P^+} 

\def\vecpf{\vec{p}_f}
\def\vecal{\vec{\alpha}}

\def\dhat{\hat{d}}
\def\dzerohat{\hat{d}_0}
\def\qone{q_1} \def\qtwo{q_2} \def\pone{p_1} \def\ptwo{p_2}
\def\qpm{q^\pm} \def\ppm{p^\pm}
\def\alplus{\alpha^+} \def\alminus{\alpha^-}
\def\alpm{\alpha^\pm}
\def\pplus{p^+} \def\pminus{p^-}
\def\qplus{q^+}  

\def\Kplus{K^+}

\def\Nhatf{\hat{N}^f}

\def\psip{\psi^+}

\def\etap{\eta^+}

\def\zetap{\zeta^+}

\def\Atil{\tilde{A}}

\def\Psizero{\Psi_0}

\def\gt{\tilde{\gamma}}

\def\chia{\chi_a}

\def\nchiepsi{:\chi\e^{-\psi}: }

\def\etap{\eta^+}

\def\Omegaket{\mid \Omega >}

\def\lm{\lambda_g}
\def\kf{K_g}

\def\xiplus{\xi^+}
\def\ximinus{\xi^-}

\def\sigmap{\sigma^+}
\def\sigmam{\sigma^-}

\def\gtil{\tilde{g}}
\def\nablah{\nabla_h}
\def\Psizero{\Psi_0}
\def\gt{\tilde{\gamma}}
\def\xplus{x^+}
\def\xminus{x^-}

\def\ff{{1\over \gamma^2}}
\def\delplus{\del_+}
\def\delminus{\del_-}
\def\xiplus{\xi^+}
\def\ximinus{\xi^-}
\def\gmn{g_{\mu\nu}}
\def\ghat{\hat{g}}

\def\etamn{\eta_{\mu\nu}}
\def\psip{{\psi_+}} \def\psim{{\psi_-}}
\def\xip{{\xi^+}}  \def\xim{{\xi^-}}
\def\xipf{{\xi^+_f}} \def\xipi{{\xi^+_i}} \def\xipz{{\xi^+_0}}
\def\sigmap{{\sigma^+}} \def\sigmam{{\sigma^-}}
\def\sigmapf{{\sigma^+_f}} \def\sigmapi{{\sigma^+_i}}
\def\zetap{{\zeta^+}} \def\zetam{{\zeta^-}}
\def\delx{\del_x} 
\def\rone{{R_1}} \def\rtwo{{R_2}}
\def\Pinv{\Phi^{-1}}
\def\Rt{\tilde{R}}
\def\nablat{\tilde{\nabla}}

\def\ginv{g^{-1}}
\def\calR{{\cal R}}

\def\papertitlepage{\baselineskip 3.5ex \thispagestyle{empty}}
\def\Title#1{\baselineskip 1cm \vspace{1.5cm}\begin{center}
 {\Large\bf #1} \end{center} 
\vspace{0.5cm}}
\def\Authors#1{\begin{center} {\it #1} \end{center}}
\def\Abstract{\vspace{1.0cm}\begin{center} {\large\bf Abstract} 
           \end{center} \parbigskip}
\def\Komabanumber#1#2{\hfill \begin{minipage}{4cm} UT-Komaba #1
              \parn #2 \end{minipage}}

\renewenvironment{thebibliography}{\pagebreak[3]\par\vspace{0.6em}
\begin{flushleft}{\Large \bf References}\end{flushleft}
\vspace{-1.0em}

\begin{enumerate}\if@twocolumn\baselineskip=0.6em\itemsep -0.2em
\else\itemsep -0.2em\fi\labelsep 0.1em}{\end{enumerate}}
\begin{document}
\papertitlepage
\vspace*{0cm}
\Komabanumber{94-18}{hep-th/9412224}
\Title{ On Quantum Black Holes
 \footnote[1]{Invited lecture given 
 at  the 13th Symposium on Theoretical Physics, Sorak, (1994),
to appear in the Proceedings.  }
} 
\parbigskipn
\Authors{{\sc 
Yoichi Kazama 
\footnote[2]{e-mail address:\quad  
kazama@hep3.c.u-tokyo.ac.jp}
}
\vskip 3ex
 Institute of Physics, University of Tokyo, \\
 Komaba, Meguro-ku, Tokyo 153 Japan \\
  }
\vspace{-0.5cm}
\parbigskipn\parmedskipn
\baselineskip=14pt
\Abstract
A pedagogical discussion is given of some aspects of  \lq\lq quantum
  black holes",  primarily using recently developed two-dimensional 
models.  After a short preliminary concerning  classical black holes, 
 we give several motivations for studying such models, especially the 
 so called dilaton gravity models in $1+1$ dimensions. Particularly 
 attractive is the one  proposed by Callan,Giddings, Harvey and
 Strominger (CGHS), which is classically solvable and contains 
 black hole solutions.  Its semi-classical as well as classical properties 
 will be reviewed, including how a flux of matter fields produces a black hole 
 with a subsequent emission of Hawking radiation. Breakdown of such 
 an approximation near the horizon, however, calls for exactly solvable 
 variants of this model and some attempts in this direction will then be 
 described. A focus will be placed on a model with 24 matter fields, 
 for which exact quantization can be performed and physical states 
 constructed.  A method will then be proposed to extract space-time 
 geometry described by these states in the sense of quantum average 
 and examples containing a black hole will be presented.  Finally 
 we give a (partial) list of future problems and discuss the nature of 
 difficulties in resolving them.   
%
\parbigskip
\newpage
\baselineskip=14pt
\section{Introduction}
\sectionnumbering
Quantum gravity is perhaps the most far-fetched, elusive, enigmatic and 
 dangerous subject in physics. When the fluctuations implied by the 
 word \lq\lq quantum" is not so large, the danger is not so acutely 
felt since we are still talking about ripples on some definite 
space-time.  But as the fluctuations become large, we begin to lose 
virtually all the familiar and useful notions in physics, as the very
 stage upon which all the events take place, namely the space-time 
 itself, melts away. Worse, we do not even know how to properly 
 quantize gravity in the background independent way.  Even if we did 
 and if we could formally solve the theory, it is still hard to 
 interpret the wave function so obtained.  \parsmallskip
Despite all this, everybody would agree that the subject must contain
 fascinating and fundamental physics, the prime example of which is 
 that of black holes.  Quantum physics of black holes is fascinating 
 because it challenges various basic notions both in quantum mechanics 
 and general relativity. In this lecture, we shall give a pedagogical 
 review of some of the recent developments concerning  \lq\lq quantum 
black holes". \parsmallskip
But, what is a \lq\lq quantum black hole"? In fact this nomenclature, 
  which often appears in the literature, does not yet have a 
 definitive meaning.  So let us try to give the answer at various 
 levels. \parsmallskip
 (i)\ At the least ambitious level, it is used to symbolize
 various effects due to {\it quantized matter fields} around a 
{\it classical black hole}. Already at this level, fascinating 
 phenomena exist and interesting questions can be asked. To cite 
 a few, the phenomena of Hawking radiation\cite{Haw(a)}
, possible loss of information
 across the event horizon and the related question of the evolution
 of a pure state into a mixed state\cite{Haw(b)} (so called the problem of 
 quantum incoherence) can all be recognized at this level, albeit 
 satisfactory answers to these questions would require understanding 
 at higher levels. Also, propagation of strings in a black hole 
background is an interesting subject in this category. \parsmallskip
(ii)\ At the next level of sophistication, one would like to include 
 at least some of the fluctuations or the deformation of the metric 
around a black hole, in response to the averaged quantum 
 effects of the matter and/or of the metric itself. 
 This is often called the problem of {\it back reaction.} \parsmallskip
(iii)\ If one follows the line of classification just described, the 
 final level would be a black hole in {\it fully quantized} gravity 
 models. One should note a profound gap between the 
 level (ii) and this level. Up to the previous level, the treatment is 
 more or less semi-classical and classically well-defined notion of
 a black hole is still clearly visible.  On the other hand, 
as will be explicitly illustrated towards the last part of this lecture,
  physically allowed  states in exactly quantizable models will be 
completely  independent of the space-time coordinates as a consequence 
 of the principle of general covariance.  Therefore, it 
 becomes quite non-trivial even to identify a black hole. Hard as it 
 may be, it is only at this level that we can hope to answer many of the
 most intriguing questions concerning black holes. Besides those 
 already mentioned at level (i), these questions include whether 
 the singularity can be washed away by quantum effects, what the end 
 point of a black hole evaporation would be like, and the statistical 
 mechanical meaning of the entropy of a black 
hole\cite{Bek}. \parsmallskip
(iv)\ Besides the three levels of increasing degree of sophistication 
just described, we must mention one more view towards the meaning of 
 a quantum black hole. It is the possibility that in full quantum 
 treatment a black hole might behave like a particle and vice 
versa\cite{bhlptcl}, \cite{HolWil}. 
 This can be motivated by recalling that given a mass scale $M$ 
 one can form two fundamental length scales, one in quantum mechanics 
 called the Compton wave length $L_{C}$ and the other in general 
relativity called the Schwarzschild radius $L_{SS}$:
\eqabegin
 L_C &=& {\hbar \over Mc} \comma  \qquad 
 L_{SS} = {GM \over c^2} \period
\eqaend
A simple yet important observation is that the Planck mass scale, 
 $M_{Pl}=\sqrt{\hbar c/G}$, is characterized non-perturbatively by that 
scale at which  $L_C$ and $L_{SS}$ are equal. This means that 
 for a black hole with \lq\lq small" mass, $M < M_{Pl}$, its Schwarzschild 
 radius is {\it inside} the Compton wave length, just like for 
 elementary particles we know. This may be relevant to the 
 problem of the fate of an evaporating black hole. 
On the other hand, if what we normally think as a particle
 has a mass $M$ bigger than $M_{Pl}$, then its horizon is located 
{\it outside} the Compton wave length, a situation typical of a black hole. 
 That  such particles naturally appear in string theory 
 makes this observation extremely interesting. In fact, evidences for 
 this phenomenon have been found in recent investigations\cite{strbkh}. 
 This may have a profound bearing on the understanding of the short
 distance behavior of the string theory. \parsmallskip
So, evidently there are many levels at which to talk about 
 \lq\lq quantum black holes". In this lecture, we shall discuss 
 aspects  (i) $\sim$ (iii) primarily using two-dimensional models, especially 
those of quantum dilaton gravity. \parsmallskip
The content of the rest of the lecture will be as follows: To make it
 more or less self-contained, we shall begin in Sec.2 with an 
 elementary review of classical black 
holes.\footnote[3]{For more detailed description
of the content of this section, see for 
example \cite{Wald}-\nocite{MTW}\cite{HE}. } 
After recalling the causal structure of flat space-time and the notion of  
Penrose diagram, we describe the basic properties of a 
 Schwarzschild black hole in several different coordinates. 
With this preliminary, we then go on in Sec.3 to give several 
 motivations for studying two-dimensional dilaton gravity models, 
 the main subjects of this lecture. One comes from the dimensional 
 reduction of the charged black hole in $3+1$ dimensions, especially 
 the so called \lq\lq extremal solution", which naturally 
 leads to the dilaton gravity model of Callan, Giddings, 
 Harvey and Strominger (CGHS)\cite{CGHS}.  Another is the recent 
 discovery that  suitable gauged Wess-Zumino-Witten (WZW) models 
 can be regarded as describing a string 
 in interaction with a black hole in two dimensional target 
space\cite{BRS},\cite{Witten},\cite{MSW}. 
 Remarks will be made on the difficulties of this approach 
 including that of incorporating the back reaction. In Sec.4, we 
 start our description of dilaton gravity models. CGHS model is 
 introduced and its classical solutions containing a black hole are 
discussed. Particularly interesting is the solution where 
 an incoming matter flux produces a black hole,  with a subsequent 
 emission of Hawking radiation. This is analyzed for 
 large number $N$ of matter fields in two different ways. Unfortunately, 
 as the mass of the black hole diminishes, this 
 \lq\lq semi-classical" approximation fails near the horizon and one is 
lead to look for exactly solvable models.  Examples of such models 
 are described in Sec.5 after discussing an inherent ambiguity in the 
 choice of models. In Sec.6, a variant of CGHS model with 
 24 massless matter fields is introduced and studied\cite{HKS}
\nocite{KS1}-\cite{KS2},\cite{VV}.  This model is 
shown to be exactly quantizable by a quantum canonical mapping into a 
set of free fields. Furthermore, by making use of conformal 
 invariance,  all the physical states can be constructed 
 in terms of free field oscillators.  However, as these states do not 
depend on space-time coordinates, it is quite non-trivial to understand
 their physical meaning. In Sec.7, we shall propose a method to 
 extract  the space-time geometry they describe in the sense 
 of quantum average\cite{KS1},\cite{KS2}. 
 It will be explicitly demonstrated that a 
 black hole geometry emerges by judicious choice of physical states. 
 Finally in Sec.8 we give a (partial) list of future problems 
 and discuss the nature of difficulties in solving them. 
\parsmallskip
The content of the first half of this lecture has a large overlap with 
 that of the excellent reviews \cite{HS},\cite{Gi}. The reader should 
 consult these references as well. 
\section{Classical Black Holes }
\sectionnumbering
\subsection{ Penrose Diagram }
What makes a black hole such a fascinating object is its peculiar 
 causal structure and, as is well-known, it can be displayed in its 
 entirety by a device called Penrose diagram. It is a way of squeezing 
 the whole universe into our hand without messing up the causal order. 
Let us briefly recall how it works when applied to the flat Minkowski 
 space in $3+1$ dimensions. Adopting the spherical coordinates, the 
 metric is given by
\eqabegin
 ds^2 &=& -dt^2+dr^2 + r^2 d\Omega^2 \comma \\
 d\Omega^2 &=& d\theta^2 + \sin^2\theta d\phi^2 \period
\eqaend
Hereafter we will often suppress the $d\Omega^2$ part. To make the 
 causal structure manifest, define the light-cone variables $u$ and 
 $v$ by
\eqabegin
 u &=& t-r  \comma\qquad 
 v = t+r \comma \\
ds^2 &=& -du\, dv \period 
\eqaend
$u$ and $v$ are of infinite range except for the 
 constraint $v \ge u$, which comes from  $v-u = 2r \ge 0$.
 Now we make a monotonic conformal transformation to define new 
light-cone  variables $( v', u' )$ of {\it finite range}:
\eqabegin
 u &=& \tan {u'\over 2}\comma \qquad v = \tan {v'\over 2} \comma \\
&&  -\pi \le u' \le v' \le \pi \period
\eqaend
Clearly, the causal property is unchanged under this mapping. 
From $u'$ and $v'$ one can define new \lq\lq time" and \lq\lq space" 
coordinates $t'$ and $r'$ as 
\eqabegin
 t' &=& \half ( v'+u' ) \comma \qquad r' = \half (v'-u') \period
\eqaend
Just like the original radial variable $r$,  $r'\ge 0$ follows from
 $v'\ge u'$.  \parsmallskip
The space-time diagram obtained by this procedure is called the 
 Penrose diagram depicted in Fig.1a: The entire space-time is 
 squeezed into a half diamond ( with of course a two-sphere attached 
 at each point inside ). Besides the preservation of the causal 
 structure, the important feature is that various infinities are 
clearly distinguished and displayed in the diagram:\parsmallskipn
$i^+$ ( future time-like infinity ) $\Longleftrightarrow$ 
$t\rightarrow \infty$ at finite$r$,  \parn
$i^-$ (past time-like infinity ) $\Longleftrightarrow$ 
$ t\rightarrow -\infty$ at finite $r$, \parn
$i^0$ ( space-like infinity ) $\Longleftrightarrow$
 $r\rightarrow \infty$ at finite $t$, \parn
$I^+$ ( future null infinity ) $\Longleftrightarrow$ 
$ v\rightarrow \infty$ at finite $u$, \parn
$I^-$ ( past null infinity ) $\Longleftrightarrow$ 
$ u\rightarrow -\infty$ at  finite $v$.
\parsmallskip
\begin{minipage}{2.5in}
\epsfxsize=5cm
\epsfysize=5cm
\epsfbox{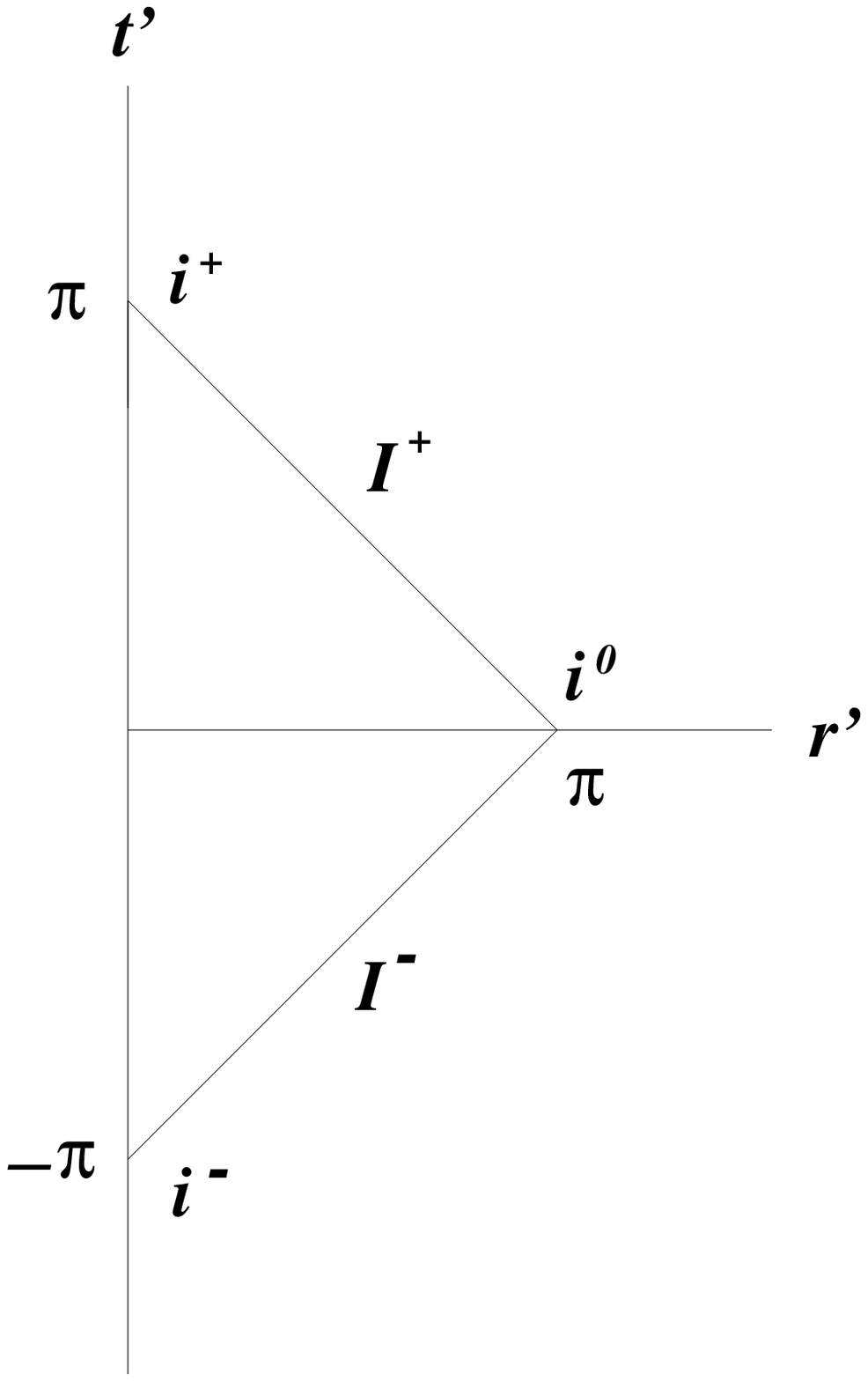}
\end{minipage}
\qquad\qquad 
\begin{minipage}{2.5in}
\epsfxsize=5cm
\epsfysize=5cm
\epsfbox{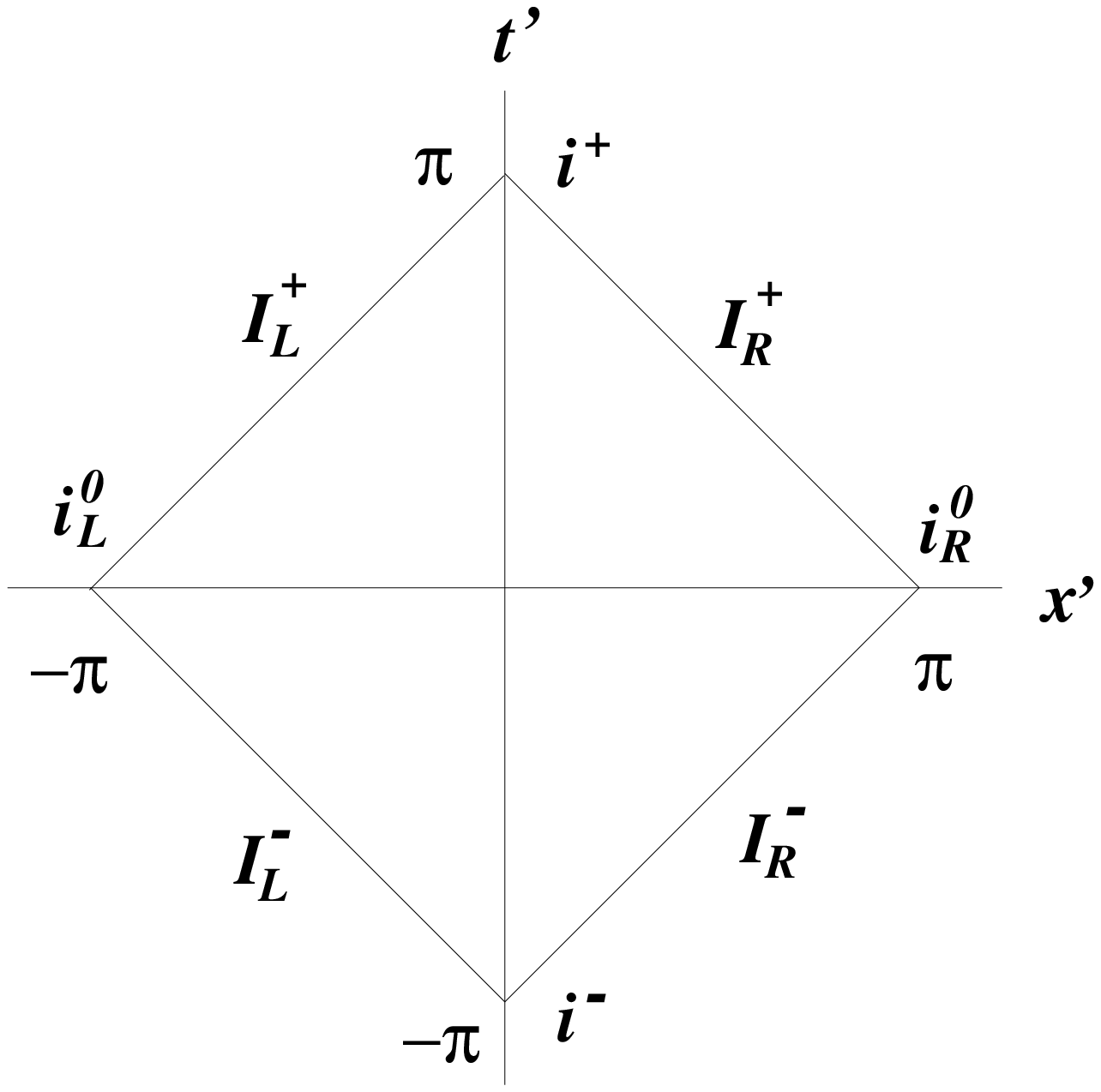}
\end{minipage}\parsmallskipn
\begin{minipage}{2.5in}
{\small{\bf Fig.1a}\quad Penrose diagram for flat Minkowski space in $3+1$ 
 dimensions.}
\end{minipage}
\qquad \qquad 
\begin{minipage}{2.5in}
{\small{\bf Fig.1b}\quad Penrose diagram for flat Minkowski space in $1+1$ 
 dimensions.}
\end{minipage}
\parbigskipn
We can follow a similar procedure for $1+1$ dimensional flat 
 space-time, relevant for our later discussion.  In this case, 
 the light-cone variables and the metric are given by 
\eqabegin
 x^\pm &=& t\pm x\comma  \qquad -\infty < x < \infty \comma \\
 ds^2 &=& -dt^2 + dx^2 = -d\xplus\, d\xminus \period
\eqaend
Because of the range of $x$, the Penrose diagram, after a conformal 
transformation, is now over the full diamond as shown in Fig.1b and 
there are pairs of space-like and null infinities. 
\subsection{ Schwarzschild Black Hole in $3+1$ dimensions }
Let us now apply the same technique to reveal the causal structure of a 
Schwarzshcild black hole. As is customary,  we start with 
its metric in the manifestly asymptotically flat 
coodinate system:
\eqabegin
 ds^2 &=& -\left(1-{2M \over r}\right)dt^2
  + {dr^2 \over \left( 1-{2M \over r}\right)} + r^2 d\Omega^2 \period
\eqaend
As $r$ passes through the horizon located at $r=2M$, the roles of 
 $t$ and $r$ are switched, implying that a light ray can get in but 
 never come out classically. 
To define  a null coordinate system,  first factor out $1-(2M/r)$ 
and define $r^\ast$ such that   
\eqabegin
 ds^2 &=& \left(1-{2M \over r}\right)\left( -dt^2 + dr^{\ast 2}\right)
\comma  \\
dr^\ast &=& {dr \over 1-{2M \over r} } \period\label{eqn:drstar}
\eqaend
To proceed, we must distinguish the exterior ($r>2M$) and the interior
 ($r<2M$) regions. In the exterior region, the solution of 
(\ref{eqn:drstar}) is given by 
\eqabegin
 r^\ast &=& r+ 2M\ln \left( {r\over 2M} -1\right) \comma 
\eqaend
which can also be expressed as 
\eqabegin
  \e^{r^\ast/2M} &=& {r \over 2M} \left( 1-{2M\over r}\right)
\e^{r/2M} \period
\eqaend
$r^\ast$ varies from $-\infty$ at $r=2M$ to $+\infty$ at $r=\infty$. 
The null coordinates, which we shall call $u^\ast$ and $v^\ast$, 
 their ranges, and the metric in terms of them are given by 
(suppressing $r^2(r^\ast)d\Omega^2$)
\eqabegin
 u^\ast &=& t-r^\ast\comma \qquad v^\ast = t+r^\ast \comma \\
 &&  -\infty < u^\ast, v^\ast < \infty \comma \\
ds^2 &=& -\left( 1-{2M\over r}\right) du^\ast\,
 dv^\ast  \period
\eqaend
Now we wish to absorb the factor in front, which makes the 
 metric singular at the horizon,  by performing a conformal 
 transformation to new null coordinates $(u,v)$, known as 
 the Kruskal coordinates.  Noting that this 
 factor occurs in $\e^{r^\ast /2M}$, we define $u$ and $v$ such that 
\eqabegin
 du\, dv &\propto & \e^{r^\ast /2M} du^\ast\, dv^\ast 
= \e^{(v^\ast-u^\ast)/4M} du^\ast\, dv^\ast \period
\eqaend
For $u$ and $v$ to be real, there are two possible pairs of signs:
\eqabegin
 du &\propto& \pm \e^{-u^\ast/4M} du^\ast \comma \\
 dv &\propto & \pm \e^{v^\ast/4M} dv^\ast \period
\eqaend
Solving them, we get two regions $(I)$ and $(II)$:
\eqabegin
&& \bracetwo{u = - \e^{-u^\ast/4M}}{v =  \e^{v^\ast/4M}\comma  } \quad (I) \\
&&  \bracetwo{u = \e^{-u^\ast/4M}}{v = - \e^{v^\ast/4M}\period } \quad (II) 
\eqaend
For both regions the metric takes the same form, which is 
 manifestly non-singular at the horizon:
\eqabegin
ds^2 &=& -{32M^3 \over r} \e^{-r/2M} dudv  \label{nsbhm} \period
\eqaend
\par
To study the interior region, we must go back to the differential 
 equation (\ref{eqn:drstar}). The appropriate real solution for 
 $r<2M$ is given by
\eqabegin
r^\ast &=& r+2M \ln \left( 1-{r\over 2M}\right) \comma 
\eqaend
where the range of $r^\ast$ is 
$-\infty\  (r=2M) < r^\ast \le 0\   (r=0) $. Define $u^\ast$ and 
$v^\ast$ exactly as before. Then we get
\eqabegin
 \e^{r^\ast /2M} &=& \e^{(v^\ast -u^\ast )/4M} = \e^{r/2M}
 \left(1-{r\over 2M}\right)  \\
 &=& -{2M\over r}  \e^{r/2M} \left( 1-{2M\over r}\right) \comma \\
 && -\infty < u^\ast \le v^\ast < \infty \comma \\
 && u^\ast =v^\ast \Longleftrightarrow  r^\ast =0 \period
\eqaend
To make the form of $ds^2$  the same 
 as for the exterior region, we require 
\eqabegin
 du\, dv &\propto & -\e^{(v^\ast -u^\ast )/4M} du^\ast dv^\ast \period
\eqaend
We then have the following two cases:
\eqabegin
&& \bracetwo{u = \e^{-u^\ast/4M}}{v =  \e^{v^\ast/4M}\comma  } \quad (III) \\
&& \bracetwo{u =- \e^{-u^\ast/4M}}{v = - \e^{v^\ast/4M}\period } 
\quad (IV) 
\eqaend
The constraint $r\ge 0$  translates into $ uv \le  1$. 
For both regions $III$ and $IV$, the line element takes the form
\eqabegin
ds^2 &=& -{32M^3 \over r} \e^{-r/2M} dudv \comma 
\eqaend
which is identical to (\ref{nsbhm}). 
The singularity at $r=0 \Leftrightarrow uv=1$ is genuine and is called 
the black hole  singularity. \parsmallskip
With the above knowledge, we can draw a $u$-$v$ diagram called 
 the Kruskal diagram ( see Fig.2).
\begin{center} 
\epsfxsize=6cm
\qquad
\epsfbox{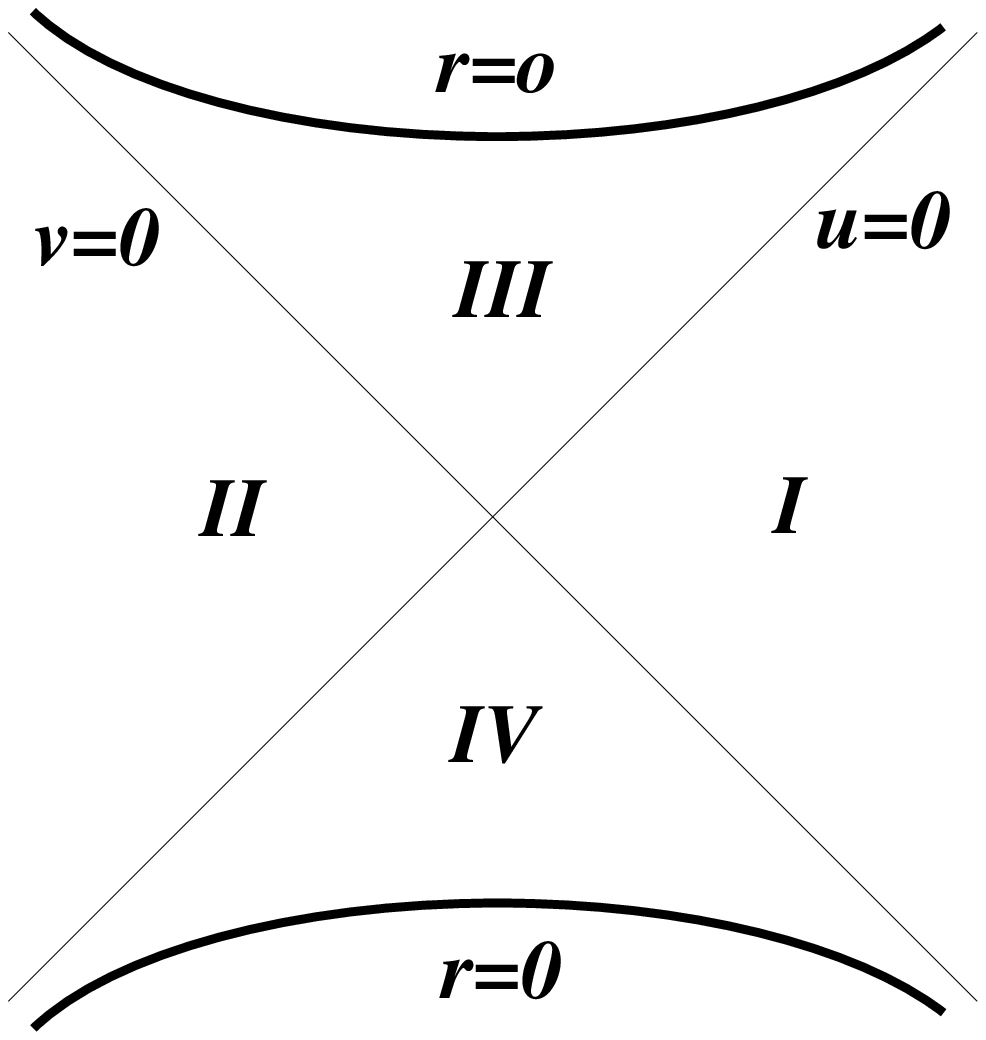}
\end{center}
\begin{center}
{\small {\bf Fig.2} \quad Kruskal diagram for  Schwarzschild black hole.}
\end{center}
\parbigskipn
It is clear that the 
 lines $u=0$ and $v=0$ separate the fate of the future-directed
 light rays (\ie whether they can escape to infinity or not) and hence 
 represent the global event horizons. \parsmallskip
The Kruskal diagram is still of infinite range. To get the Penrose 
 diagram, we drop the front factor of the metric, which does not 
 change the causal structure, and make the now familiar conformal
 transformation to new null coordinates $(u', v') $:
\eqabegin
 u &=& \tan{u' \over 2} \comma \qquad v = \tan {v' \over 2} \period
\eqaend
By using the simple identity
\eqabegin
 \tan^{-1}\left( {u+v \over 1-uv}\right) &=& \tan^{-1}u +\tan^{-1}v 
 = \half (u'+v') \comma 
\eqaend
we easily see that the singular line at $r=0 \ (uv=1)$ are mapped 
 into the horizontal straight lines $u'+v'=\pm \pi$. The 
 resultant Penrose diagram is depicted in Fig.3. 
\begin{center} 
\epsfxsize=8cm
\quad\epsfbox{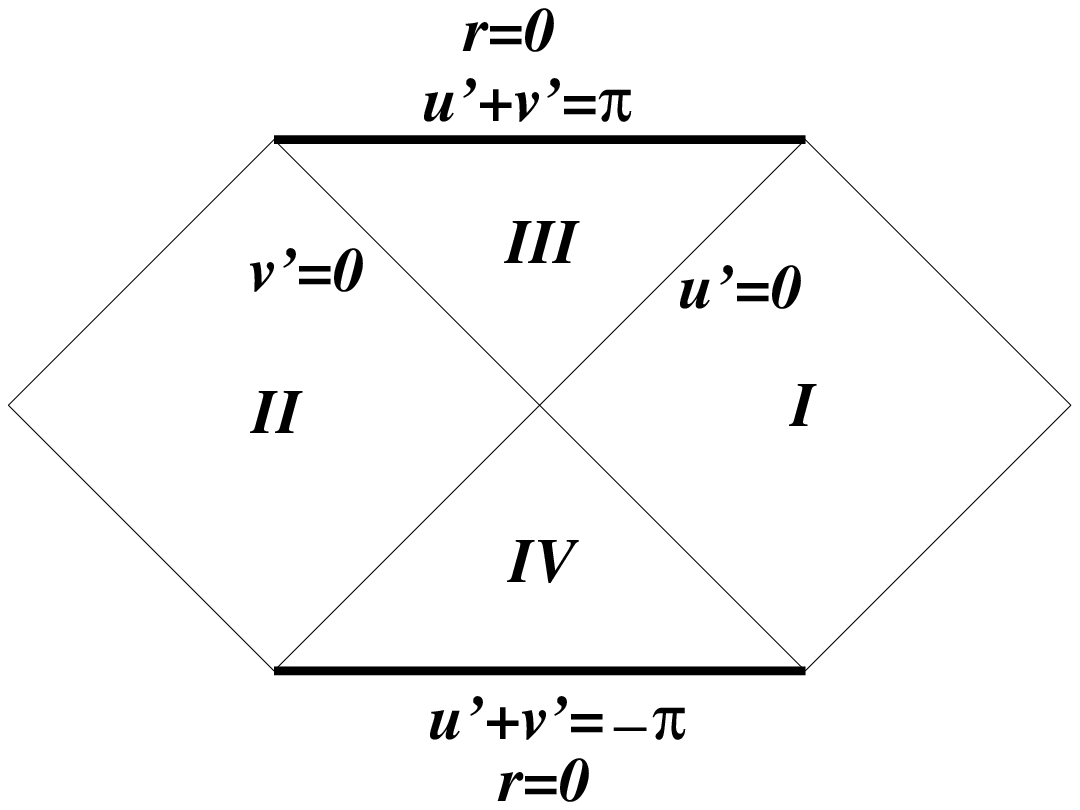}
\end{center}
\begin{center}
{\small {\bf Fig.3} \quad Penrose diagram for  Schwarzschild black hole.}
\end{center}
\parbigskipn
For later comparison, it is of interest to look at the form of the 
 singularity in the Kruskal coordinates. By expressing the factor
 $(1/r)\e^{-r/2M}$ in terms of $u$ and $v$, we find the square root 
 type behavior 
\eqabegin
 ds^2 &\simeq & -8\sqrt{2} M ^2{1\over \sqrt{1-uv}} du dv \period
\eqaend
Note that the strength of the singularity  is not expressed in the 
Penrose diagram. 
\section{ Motivation for 2 Dimensional Dilaton Gravity Models }
\sectionnumbering
\subsection{ Black Hole in $3+1$ Dimensional  Dilaton Gravity and 
Dimensional Reduction} 

One of the characteristic features of (super)string theory is that 
 a scalar field called dilaton is always associated with the graviton 
and  plays an important role in gravitational physics.  As it is 
 massless (at least perturbatively),  it remains in the low energy 
effective action after suitable compactification down to $3+1$ 
dimensions.  The simplest among such actions contains, in addition, 
 a $U(1)$ gauge field and takes the following form:
\eqabegin
 S &=& \int dx^4 \sqrt{-g}\, \e^{-2\phi} \left( R + 4g^{\mu\nu}
 \nabla_\mu\phi \nabla_\nu\phi -\half g^{\mu\lambda}g^{\nu\rho}
 F_{\mu\nu} F_{\lambda\rho} \right) \period
\eqaend
Although this is the natural form following from string theory ( for
 example by $\beta$ function analysis of the corresponding non-linear
 sigma model), it apparently has two \lq\lq unusual" features. First, 
 with the factor $\e^{-2\phi}$, which plays the role of the string coupling
 constant squared $g_s^2$,  the Einstein term does not have 
 the canonical form. Furthermore,  the kinetic term for the 
 dilaton  has the \lq\lq wrong" sign. (We are using the space-favored
 signature convention.) These features are not necessarily problematic
 since there is always an (admittedly annoying) ambiguity in the 
 choice of the conformal frame. In fact, if we make a conformal 
 transformation of the form
\eqabegin 
 \tilde{g}_{\mu\nu} &=& \e^{-2\phi} g_{\mu\nu} \comma
\eqaend
both of the above \lq\lq problems" are removed simultaneously and 
 we get the canonical form:
\eqabegin
 S &=& \int dx^4 \sqrt{-\gtil}\left( \tilde{R} -2\gtil^{\mu\nu}
 \nabla_\mu\phi \nabla_\nu\phi - \half \e^{-2\phi}
 \gtil^{\mu\lambda}\gtil^{\nu\rho}
 F_{\mu\nu} F_{\lambda\rho} \right) \period
\eqaend
We shall follow the usual terminology and call $\gtil_{\mu\nu}$ the 
 \lq\lq canonical"  metric and $g_{\mu\nu}$ the \lq\lq string metric". 
In this model, there is no potential for the dilaton field. This means 
 that one can choose any constant value $\phi_0$ for it at infinity, 
 which therefore is a parameter of the model. This parameter has a 
 definite physical meaning in the canonical frame: The exponential 
 factor $\e^{2\phi_0}$, which appears in front of the gauge field 
kinetic term, can be identified as  the  electromagnetic coupling squared 
at $\infty$. \parsmallskip
Although this model has a variety of black hole solutions\cite{GM}, 
\cite{GHS},
 we shall focus on the simplest non-rotating one with a magnetic charge 
$Q$. Its line element in the canonical frame is given by
\eqabegin
d\tilde{s}^2 &=& \gt_{\mu\nu}dx^\mu dx^\nu \nn\\
 &=& -\left( 1-{2M\over r}\right) dt^2 + {dr^2\over ( 1-2M/r)}\nn\\
 && \qquad + r^2\left( 1-{Q^2\over 2Mr}\e^{-2\phi_0}\right) d\Omega^2 
 \comma \\
\e^{-2\phi} &=& \e^{-2\phi_0}\left( 1-{Q^2\over 2Mr}\e^{-2\phi_0}\right)
 \comma \\
 F &=& Q\sin\theta d\theta \wedge d\phi \period
\eqaend
Notice that the gauge field configuration is precisely that of an 
 abelian monopole. The metric is quite similar to that of Schwarzschild
 solution except that the area of the spatial two-sphere is proportional
 to the factor $\left( 1-{Q^2\over 2Mr}\e^{-2\phi_0}\right)$, which 
 also appears in $\e^{-2\phi}$. One can check that a curvature singularity
 develops as this factor vanishes.  It is also easy to see that the 
 horizon is located at $r=2M$ just like in the Schwarzschild case. 
 \parsmallskip
What will be relevant to $1+1$ dimensional dilaton gravity model is 
 a special case of the above solution called the \lq\lq extremal 
solution". It occurs at a specific value of the magnetic charge, 
 namely at $Q^2 = 4M^2\e^{2\phi_0}$, where 
 the factor for $\e^{-2\phi}$ becomes exactly the one specifying
 the horizon:
\eqabegin
 \e^{-2\phi} &=& \e^{-2\phi_0}\left( 1-{2M\over r}\right) \period
\eqaend
As seen in the canonical frame, 
this is a peculiar limit where the singularity and the horizon coincide
 and at the same time the area of the two-sphere vanishes. In the 
 string metric, however, the situation looks quite different. As 
 we have to mulitiply by $\e^{2\phi}$ to go to the string 
 metric, the singularity in the angular part and in front of 
 $dt^2$ dissappear simultaneously and we get 
\eqabegin
 ds^2 &=& \e^{2\phi_0}\left( -dt^2+{dr^2\over ( 1-(2M/r))^2}
 + r^2d\Omega^2 \right)\period
\eqaend
The asymptotic behavior for large $r$ is unchanged and we still
 have a flat space. To investigate what happens near $r=2M$ 
let us introduce the Kruskal coordinate $\sigma$ (previously 
 denoted by  $r^\ast$): 
\eqabegin
 d\sigma &=& {dr \over 1-2M/r} \comma \\
 \e^{\sigma/2M} &=& {r\over 2M}\e^{r/2M} \left( 1-{r\over 2M} \right) 
\period
\eqaend
The metric and the dilaton field then take the forms 
\eqabegin
 ds^2 &=& \e^{-2\phi_0}\left( -dt^2+ d\sigma^2 + r^2(\sigma)d\Omega^2
 \right) \comma \\
\e^{-2\phi} &=& \e^{-2\phi_0}\e^{\sigma/2M}\, {2M\over r}\e^{-r/2M}
\label{eqn:etwophi}\period
\eqaend
Now it is easy to see what happens near $r=2M$. We get 
\eqabegin
ds^2 &\longrightarrow & \e^{2\phi_0} \left( -dt^2 + d\sigma^2
 + (2M)^2 d\Omega^2 \right) \comma \label{metnearhor}\\
\phi &\longrightarrow & \phi_0-{\sigma\over 2M} \period
\eqaend
Eq.(\ref{metnearhor}) says that 
we have a two-sphere of fixed radius $2M$ and the space-time is flat. 
As for the dilaton, it becomes linear in $\sigma$, where $\sigma 
 \rightarrow -\infty$  as seen from (\ref{eqn:etwophi}). 
This configuration is often  referred to as the {\it linear dilaton vacuum}.  
Since the two-sphere is effectively frozen, this suggests that 
 near $r=2M$ the essence of the dynamics should be 
 describable by a two-dimensional theory.  Employing the technique of 
 dimensional reduction, one obtains a simple action:
\eqabegin
 S &=& {1\over \gamma^2} \int d\xi^2 \sqrt{-g}\, \e^{-2\phi}
 \left( R + 4(\nabla \phi)^2 + 4\lambda^2 -\half F^2\right) \comma \\
\lambda^2 &=& {1\over 4Q^2} \period
\eqaend
Furthermore, if there are no charged particles, we can ignore the 
 two-dimensional gauge fields since there will be no local dynamics. 
Thus we arrive at a simple model of two-dimensional dilaton gravity
 proposed by Callan, Giddings, Harvey and Strominger\cite{CGHS}.
 We shall  describe its classical and quantum properties in detail in Sec.4.
\subsection{ $SL(2)/U(1)$ Black Hole in $1+1$ Dimensions}
Another motivation for studying $1+1$ dimensional theories comes 
 from the discovery\cite{BRS}, \cite{Witten}, \cite{MSW}  that 
a suitable gauged WZW model, the prototype of which is based on 
 the coset $SL(2,R)/U(1)$, can be regarded as an 
 exactly  conformally invariant non-linear sigma model describing a 
string theory in a black hole background. As it is the first 
explicit model capable of describing the interaction of a string 
 with a black hole, vigorous investigations have been 
 performed.
 However, as we shall discuss in a moment, this approach 
 has a couple of serious shortcomings, including the inherent 
 inability to take into account the back reaction of the metric. 
In this respect, the field theoretic model introduced at the end 
 of the previous subsection would be more versatile.  For this 
 reason, we shall keep the exposition very brief. \parsmallskip
Let us begin with the description of the ungauged WZW model based on 
$SL(2,R)$. Its action is given by 
\eqabegin
 S(g) &=& L(g) + i\Gamma(g) \comma \\
 L(g) &=& {k \over 8\pi} \int_\Sigma \sqrt{h}\, 
 h^{ij} \trace \left( \ginv \del_i g \ginv \del_j g\right) \comma \\
 \Gamma(g) &=& {k\over 12\pi} \int_B \trace \left( \ginv dg \wedge
 \ginv dg \wedge \ginv dg \right) \comma 
\eqaend
where $k$ is so called the level, which plays the role of
 the inverse coupling constant. The model has a global symmetry 
group $SL(2,R)_L \otimes SL(2,R)_R$ acting on $g\in SL(2,R)$ 
as $AgB^{-1}$, with $A\in SL(2,R)_L$ and $B \in  SL(2,R)_R$.
\parsmallskip
As the group is non-compact, there is a unitarity problem as it stands.
However, this can be cured if one can gauge away  the unwanted states. 
One way of achieving this is to gauge the anomaly-free $U(1)$ 
 subgroup generated by $\sigma_3$. Explicitly, 
\eqabegin
 A &=& B= 1 + \epsilon\, \sigma_3 + \cdots \comma \\
  \delta g &=& \epsilon \left\{ \matrixii{1}{0}{0}{-1} g
 + g \matrixii{1}{0}{0}{-1} \right\} \comma 
\eqaend
where $\epsilon$ is an infinitesimal  gauge function. 
A convenient parametrization for $g$ is 
\eqabegin
 g &=& \matrixii{a}{u}{-v}{b}  \comma \qquad ab+uv =1 \period
\eqaend
After introducing the $U(1)$ gauge field $A_\mu$ such that 
$\delta A_\mu = -\del_\mu \epsilon$, one must fix a gauge.  
  From the form of the $U(1)$ generator, it is clear 
 that demanding a relation between $a$ and $b$ does the job. 
 The simplest choice which respects the constraint $ab + uv =1$ 
 is to take $a=b$ for $1-uv>0 $ and $a=-b$ for $1-uv<0 $.  
 The condition for conformal invariance reads
$c=(3k/(k-2)) -1 =26$, where $-1$ is due to the gauging of $U(1)$. Hence
 the level should be taken to be $k=9/4$.  Upon performing the 
 integration over $A_\mu$,  the action takes 
 the form of a non-linear sigma model
\eqabegin
 L &=& -{k\over 4\pi} \int d^2x \sqrt{-h}\, 
{h^{\mu\nu}\del_\mu u \del_\nu v \over 1-uv} 
\comma  
\eqaend
from this one can immediately read off the target space line 
 element:
\eqabegin
 ds^2 &=& -{du\, dv \over 1-uv} \period 
\eqaend
This evidently  describes a black hole in a Kruskal-like coordinate. 
(As we shall see, this form is essentially identical to the black
 hole solution in the dilaton gravity model of CGHS.) 
Actually, proper treatment of the integration measure for $A_\mu$
 generates a target space dilaton field coupled to the two-dimensional
 curvature scalar $R$.  One can also introduce  the tachyon field 
 and study the Hawking radiation\cite{DijVer}. \parsmallskip
Now we would like to make some remarks on the serious shortcommings 
of the sigma model formulation of string theory in curved background
 in general. 
\parsmallskip
(i)\ The first unsatisfactory feature is the apparent redundancy in 
the  description of the metric degrees of freedom.  While the 
  fluctuations of the metric corresponding to gravitons are described 
 as excitations of a string, the macroscopic part of the metric 
  appears explicitly in the action and is regarded as 
 describing the space-time in which the very string propagates. 
 Since coherent states of gravitons should be able to represent the 
 macroscopic space-time as well, there must at least be some 
 consistency condition between the two. In other
 words, the correct formulation must be invariant under 
 the choice of decomposition of the background and the fluctuation. 
\parsmallskip
(ii)\ We may illuminate the same kind of problem in a different way. 
Normally, to incorporate the fluctuations of a background field one 
 trys to integrate over the relevant field, in this case the 
 target space metric $G_{\mu\nu}$. But since only for a special class of 
$G_{\mu\nu}$ is the model conformally invariant, integration over 
 all possible metric will inevitably 
require some off-shell formulation of string theory, which is outside the
 scope of the sigma model description. String field theory would be the 
 prime candidate but its present day formulation does not appear 
 to be suitable for black hole physics.  Another possibility is to 
 insist on the conformal invariance and integrate only over 
 the  moduli space with some measure.  The problem here is that 
 there does not appear to be any principle of determining the 
 appropriate measure. 
\parsmallskip
(iii)\ It should by now be clear that the gist of the problem 
 is the lack of background independent formulation of string theory. 
 Manifestation of this can be seen already at the level of 
 low energy effective field theory: If one can identify 
 the moduli parameters as expectation values of some fields, one would
 have a background independent formulation.  However, 
 the potential for these fields will be  flat and one cannot 
 determine the preferred direction. \parsmallskip
With the above discussion, it is fair to say that at present there
 is no background independent formulation of string theory in which 
 one can discuss quantum black holes. One can only describe a black hole
 as a rigid background geometry and hence cannot address the 
 question of back reaction. ( A possible exception may be the attempts 
 to describe black hole geometry in suitable matrix models\cite{JY}. 
 It is interesting to see if one can overcome the difficult problem 
 of space-time interpretation in that approach. )
\section{Semi-Classical Black Holes in $1+1$ Dimensional
 Dilaton Gravity Models }
\sectionnumbering
With the motivations just described, we shall now begin our discussion
 of two-dimensional dialton gravity models of CGHS type. 
\subsection{CGHS Model}
The action of the original CGHS model is given by\cite{CGHS}
\eqabegin
 S &=& \ff \int d^2\xi \sqrt{-g}\, \Biggl\{ \e^{-2\phi}\left[R+4(\nabla 
\phi)^2 + 4\lambda^2 \right] 
 - {1\over 2}\sum_{i=1}^N (\nabla f_i)^2 \Biggr\} \comma 
\eqaend
where $f_i\ (i=1,2, \ldots, N)$ are massless matter scalar fields. 
Let us take the conformal gauge with flat background and use the 
 light-cone variables. Relevant expressions are 
\eqabegin
 \gmn &=& \e^{2\rho}\eta_{\mu\nu} \comma \\
\xi^\pm &=& \xi^0 \pm \xi^1 \comma \qquad  
\Box = \eta^{\mu\nu}\del_\mu\del_\nu =-4\delplus\delminus\comma \\
 g_{+-} &=& -{1\over 2}\e^{2\rho} \comma \qquad g^{+-} = -2 \e^{-2\rho}
 \comma \\
R_{\mu\nu} &=& -\etamn \Box \rho\comma \qquad 
R_{+-} = -2\delplus\delminus\rho \comma \\
R &=& -2\e^{-2\rho}\Box\rho = 8\e^{-2\rho}\delplus\delminus \rho \period
\eqaend
Then the action becomes
\eqabegin
S &=&{4\over \gamma^2}\int d^2\xi \Biggl\{ \e^{-2\phi}
\left[ 2\delplus\delminus \rho
 -4\delplus\phi\delminus\phi + \lambda^2\e^{2\rho}\right] 
  + \half\sum \delplus f_i \delminus f_i \Biggr\} \period
\eqaend
As far as the classical analysis is concerned, it is 
 made extremely simple by making a change of 
 variable by a conformal transformation to the \lq\lq canonical metric"
 $\gtil_{\mu\nu}$ discussed in the previous section. Defining 
\eqabegin
 \Phi &\equiv & \e^{-2\phi} \comma 
\eqaend
the transformation is expressed as 
\eqabegin
 g_{\mu\nu} &=& \Pinv \gtil_{\mu\nu} \comma \\
\sqrt{-g} &=& \Pinv \sqrt{-\gtil} \comma \label{eqn:sqrtg} \\
R &=& \Phi \left(  \Rt - 2\gtil^{\mu\nu}\nablat_\mu \nablat_\nu 
\phi\right)\period \label{eqn:Rtransf}
\eqaend
Notice that in two-dimensions the transformation factor for 
 $\sqrt{-g}$ in (\ref{eqn:sqrtg}) is $\Pinv$ and not $\Phi^{-2}$ 
 as in four-dimensions. Because of this, the term involving the 
 scalar curvature will not be of Einstein form ( which is 
  a total derivative in two-dimensions). Also, from (\ref{eqn:Rtransf})
 one immediately sees that the kinetic terms for $\phi$ cancel. 
The resultant action is 
\eqabegin
 S &=& \ff \int d^2\xi \sqrt{-\gtil}\, \left( \Phi \Rt 
+ 4\lambda^2-\half \sum(\tilde{\nabla} f_i)^2\right) \period
\eqaend
Now we choose a conformal gauge with the conformal factor 
 denoted by $\psi$:
\eqabegin
\gtil_{\mu\nu} &=& \e^{\psi} \eta_{\mu\nu} \comma \\
 S&=&
 {4\over \gamma^2} \int d^2\xi \left( -\half (\delplus\Phi
\delminus\psi +\delplus
\psi\delminus\Phi) + \lambda^2 \e^{\psi} \right. \nn\\
&& \left. \quad + \half \sum \delplus f_i\delminus f_i 
\right)   \period
\eqaend
The equations of motion following from the variations $\delta \psi, 
 \delta \Phi$ and $\delta f_i$ are quite simple:
\eqabegin
\delta \psi:\quad && \delplus\delminus \Phi+\lambda^2 \e^\psi =0
\comma \label{eqn:psieq} \\
\delta \Phi: \quad && \delplus\delminus\psi =0 \comma \\
\delta f_i: \quad && \delplus\delminus f_i =0 \period
\eqaend
The second equation tells us that $\psi$ is a free field and can 
 be decomposed into the right- and left-going parts:
\eqabegin
\psi &=&\psip(\xip)+\psim(\xim) \period
\eqaend
To solve the first equation (\ref{eqn:psieq}), 
define two functions $A(\xiplus)$ and $B(\ximinus)$ such that 
\eqabegin
 \delplus A &=& \lambda \e^\psip \comma \\
\delminus B &=& \lambda \e^\psim \period
\eqaend
Then (\ref{eqn:psieq}) takes the form 
\eqabegin
\delplus\delminus\left(\Phi +AB\right) &=& 0\comma 
\eqaend
showing that the combination $\Phi+ AB$ is again a free field, which 
 we call $\chi$. Thus  we have
\eqabegin
 \Phi &=& -\left(\chi +AB\right) \comma \\
\chi &=& \chi_+(\xip) + \chi_-(\xim)  \period
\eqaend
\indent 
In addition to the equations of motion, we must impose the 
 vanishing of the energy-momentum tensor, which follows from general 
 covariance. First, $T_{+-}$, which is proportional to the trace 
 $T_\mu^\mu$ for flat background we are considering, is given by 
\eqabegin
T_{+-}&=& -\left(\delplus\delminus\Phi +\lambda^2\e^\psi\right)\comma 
\eqaend
where we have set $\gamma^2 = 4\pi$ for convenience. 
This automatically vanishes from the equation of motion 
(\ref{eqn:psieq}) showing that the system has conformal invariance. 
The remainig components  $T_{\pm\pm}$ take the form 
\eqabegin
T_{\pm\pm} &=&- \del_\pm\Phi\del_\pm\psi +\del_\pm^2\Phi 
 + \half \sum (\del_\pm f_i)^2 \\
&=& \del_\pm \chi \del_\pm \psi -\del_\pm^2\chi + T^f_{\pm\pm}
\comma \label{eqn:Tpmpm}
\eqaend
where the expression in terms of the free fields $\psi$ and $\chi$ 
 (the second line) is obtained by the substitution 
$ \Phi =-\left(\chi +AB\right) $.  To solve $T_{++}=0$, 
let us note the identity 
\eqabegin
\delplus\left(\delplus\chi \e^{-\psip}\right) &=-&\left(
\delplus\chi\delplus\psi -\delplus^2\chi \right)\e^{-\psip} \period
\eqaend
Thus  $T_{++}=0$ reads $\delplus\left(\delplus\chi \e^{-\psip}\right) 
=\e^{-\psip}T^f_{++}$ and it can be integrated to yield 
\eqabegin
 \chi_+ &=& const. +\int^{\xi^+} du \e^{\psip(u)}
\int^{u}dv \e^{-\psip(v)}\,
 T^f_{++}(v) \period
\eqaend
Solution for $\chi_-$ is entirely similar. 
\parsmallskip
Before we go on, it would be useful to summarize what we got so
 far: (For simplicity we shall hereafter use $\lambda$ and 
 $1/\lambda$ as our mass and length units respectively and set 
 $\lambda=1$.)
\eqabegin
g^{\mu\nu} &=& \e^{-2\rho}\eta^{\mu\nu} =\Phi\e^{-\psi}\eta^{\mu\nu}
\comma \label{gmunu} \\
\psi &=&=2(\rho -\phi) =\psip(\xip)+\psim(\xim) =\mbox{ free field} \comma \\ 
\Phi &=& \e^{-2\phi} = -\left(\chi +AB\right) \comma \\
\chi &=& \chi_+(\xip) + \chi_-(\xim) =\mbox{ free field}\comma \\
\delplus A &=& \e^{\psip(\xip)}\comma \qquad 
\delminus B = \e^{\psim(\xim)}\comma  \label{ABeq}\\
T_{\pm\pm} &=& \del_\pm \chi \del_\pm \psi -\del_\pm^2\chi + 
\half \sum (\del_\pm f_i)^2 \comma \\
\chi_+ &=& const. +\int^{\xi^+} du \e^{\psip(u)}
\int^{u}dv \e^{-\psip(v)}\,
 T^f_{++}(v) \period \label{chiitoT}
\eqaend
\parsmallskip
Now let us describe some interesting classical solutions of this 
 system. Analysis is the simplest in the gauge (within the 
 conformal gauge) where $\psi=0$ \ie 
 $\rho=\phi$, which we shall call the Kruskal gauge. The reason why 
 we can impose such a condition is because  $\rho$, and hence $\psi$, 
transforms inhomogeneously under conformal tranformation. 
\parsmallskip
First, let us consider the case where matter fields are absent \ie 
  $f_i=0$ for all $i$ and hence $T^f_{\pm\pm}=0$. Taking into account 
 the gauge condition,  we readily obtain from (\ref{ABeq}) and 
 (\ref{chiitoT}) 
\eqabegin
\chi &=& const. \equiv -M \comma \\
A&=&\xip\comma \quad B=\xim\quad (\mbox{up to a constant shift})\period
\eqaend
Therefore,  $\rho$, $\phi$, the line element and the curvature 
 scalar become 
\eqabegin
\e^{-2\rho} &=&\e^{-2\phi}= M-\xip\xim\comma  \\
 ds^2 &=&- { d\xi^+ d\xi^- \over  M -\xi^+\xi^-}\comma \\
R &=& 8 \e^{-2\rho}\delplus\delminus\rho 
= {4 M \over  M -\xi^+\xi^-} \period
\eqaend
This is similar to the Schwarzschild solution  in the Kruskal 
coordinate and in fact exactly the same as the 
$SL(2)/U(1)$ black hole, with $M$ as the mass of the black hole. 
 In Fig.4  we show the Penrose diagram 
 for this configuration.
\begin{center} 
\epsfxsize=8cm
\quad\epsfbox{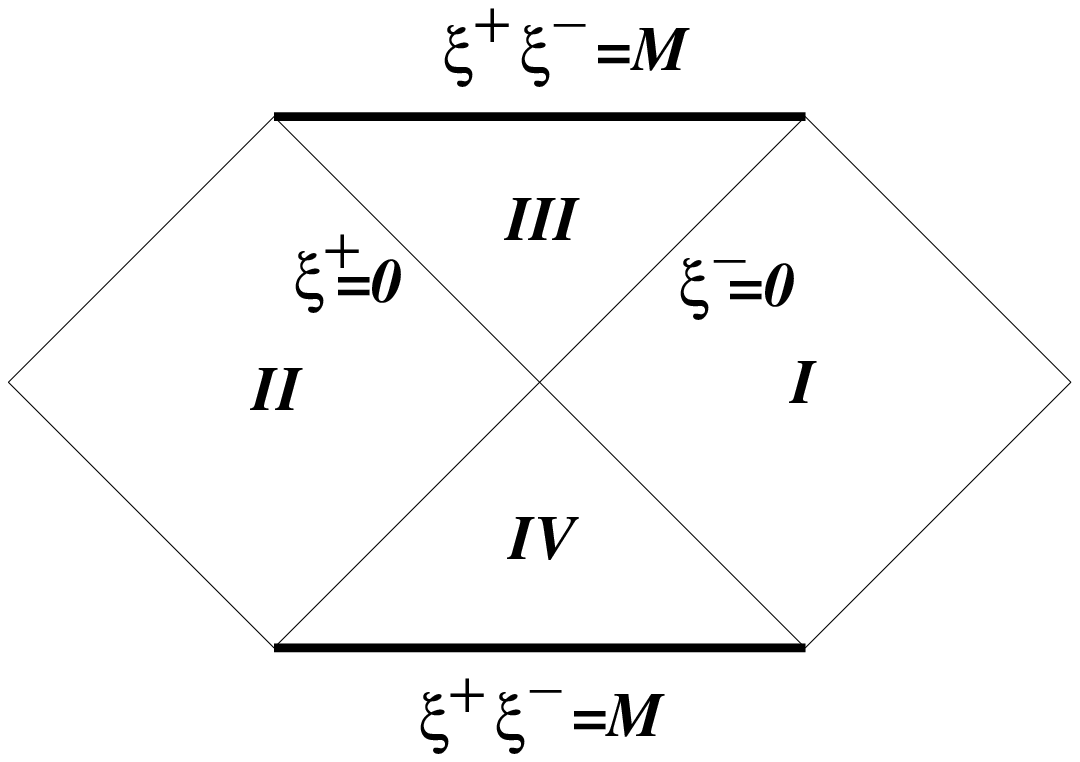}
\end{center}
\begin{center}
{\small {\bf Fig.4} \quad Penrose diagram for a black hole in CGHS
 model.}
\end{center}
\parbigskipn
 Note the horizons at $\xip=0$ and $\xim=0$
 and the curvature singularity along $\xip\xim=M$. 
\parsmallskip
From the expression of $R$ we see that as $|\xiplus\ximinus|
 \rightarrow \infty$ the space-time becomes flat. The coordinates 
 in which this becomes manifest can easily be constructed by a 
 conformal transformation. In region $I$ ($\xi^+>0\comma 
\xi^-<0$) the tranformation, which is similar to the one between 
 the Kruskal and $(u^\ast, v^\ast)$ coordinates in the Schwarzschild 
 case, is given by 
\eqabegin
 \xi^+ &=& \e^{\sigma^+}\comma \qquad d\xip =\xip d\sigmap \comma \\
  \xi^- &=&- \e^{-\sigma^-} \comma\qquad d\xim=-\xim d\sigmam \period
\eqaend
The line element then becomes
\eqabegin
  ds^2 &=& -\e^{2\rho(\sigma)} d\sigma^+ d\sigma^- \nn\\
&=& -{d\sigma^+ d\sigma^- \over 1 + M
 \e^{-(\sigma^+-\sigma^-)}}\nn\\
&\stackrel{\large\sigma\rightarrow\infty}{\longrightarrow}& 
-d\sigmap d\sigmam  \comma 
\eqaend
where $\sigma$ is the spatial coordinate defined by $2\sigma \equiv 
 \sigma^+-\sigma^-$. Notice that $\rho$, not being a scalar, 
 got non-trivially transformed. In contrast, $\phi$ is a genuine scalar 
 and we still have $\e^{-2\phi}=M+\xiplus\ximinus=
M + \e^{\sigma^+ -\sigma^-}$. In the asymptotically flat region
 it behaves like 
\eqabegin
 \phi(\sigma) &=& -\sigma -\half \ln\left(
 1+M\e^{-2 \sigma} \right)\nn\\
 &\stackrel{\large\sigma\rightarrow\infty}{\longrightarrow} &
 - \sigma \comma 
\eqaend
and we have  a  linear dilaton vacuum. \parsmallskip
Let us recall that $\e^{2\phi}$ carries the meaning of  $g_s^2$, 
the string  coupling constant squared. It represents the strength 
 of the joining-splitting interaction of strings in the original theory
 and in the present context indicates the strength of the gravitational 
 coupling. From the expression above it is easy to see that (reinstating 
 the scale $\lambda$) 
\eqabegin
 g^2_s &\sim & {\lambda \over M} \qquad
 \mbox{ near the horizon} \comma \label{gsnearhor}\\
         &\sim & 0 \qquad\quad \mbox{in the linear dilaton region}
\period \nn
\eqaend
We see that near the horizon the coupling becomes large as 
 $M$ becomes small.  \parsmallskip
So much for the case without matter fields. Now we shall discuss a 
 more interesting solution which describes a formation of a black 
 hole by a flux of matter fields. For definiteness, suppose 
 a pulse of left-going  matter flux is sent in between 
 $\xi^+_i$ and $\xi^+_f$. It is represented by the matter 
energy-momentum tensor of the form 
\eqabegin
 T^f_{++}(v) &=& F(v)( \theta(v-\xi^+_i) -\theta(v-\xi^+_f)) \comma 
\eqaend
where $F(v)$ is a positive function specifying the profile of the flux
 (see Fig.5). 
\begin{center} 
\epsfxsize=8cm
\quad\epsfbox{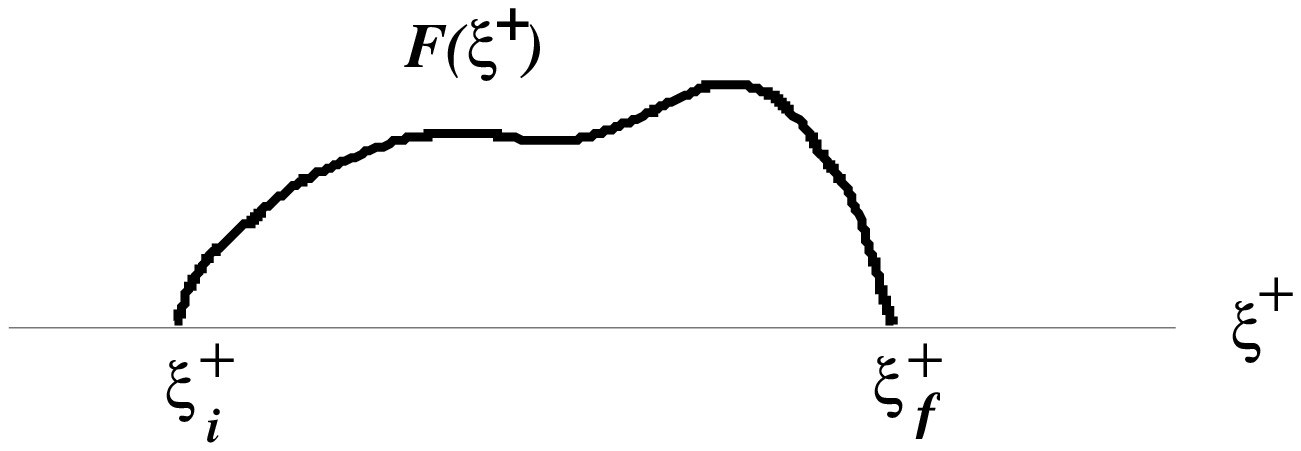}
\end{center}
\begin{center}
{\small {\bf Fig.5} \quad Profile of left-going matter flux.}
\end{center}
\parbigskipn
Assume we have a linear dilaton vacuum before the matter is sent in. 
Then, from the general formula (\ref{chiitoT}), $\chi$ is given by 
\eqabegin
 \chi &=& \theta(\xi^+-\xi^+_i)\int^{\xi^+}_{\xi^+_i} du 
\int^u_{\xi^+_i} dv T^f_{++}(v) \nn\\
 &=& \bracebegin{lll}
 0 & \qquad \xip \le \xipi \\
H(\xip) & \qquad \xipi \le \xip \le \xipf \\
(\xip -\xipf)a +b & \qquad \xipf\le \xip 
\braceend 
\comma 
\eqaend
where
\eqabegin
 H(u) &\equiv &  \int^u_{\xi^+_i} dw\int^w_{\xi^+_i} dv F(v)\comma \\
H'(u) &=& \int^u_{\xi^+_i} dv F(v) \comma \\
b &=& H(\xipf)\comma\qquad a=H'(\xipf) \period
\eqaend
Thus {\it after the matter has collapsed in}, the metric is of the form
\eqabegin
 \e^{-2\rho} &=& \e^{-2\phi} = -(\chi + AB) \nn\\
&=& (\xipf a -b ) -\xip(\xim+a ) \nn\\
 &=& M -\xip(\xim+a ) \period  
\eqaend
 $M$ is defined as $M \equiv  \xipf a-b $ and it can be shown to
 be positive for any positive $F(v)$. \parsmallskip
The behavior of the scalar curvature can be readily computed. 
It is given by 
\eqabegin
R &=& \bracebegin{lll}
 0 &\qquad \xip \le \xipi \\
-{4\left(\xip H'(\xip) -H(\xip)\right)
 \over H(\xip)+\xip\xim } & \qquad \xipi \le \xip \le \xipf \\
{4M \over M-\xip(\xim +a) }& \qquad \xipf\le \xip 
\braceend \period
\eqaend
Before the matter flux is sent in, $R$ naturally vanishes. After 
 the flux has past, it takes 
 exactly the same form as for the previously discussed 
 black hole solution without matter, except 
 with an important shift $\ximinus \rightarrow \ximinus + a$ 
 caused by  the flux. The behavior in the transient region is a smooth 
 interpolation of these two. The Penrose diagram of this space-time 
 is shown in Fig.6. \parsmallskip
\begin{center} 
\epsfxsize=8cm
\quad\epsfbox{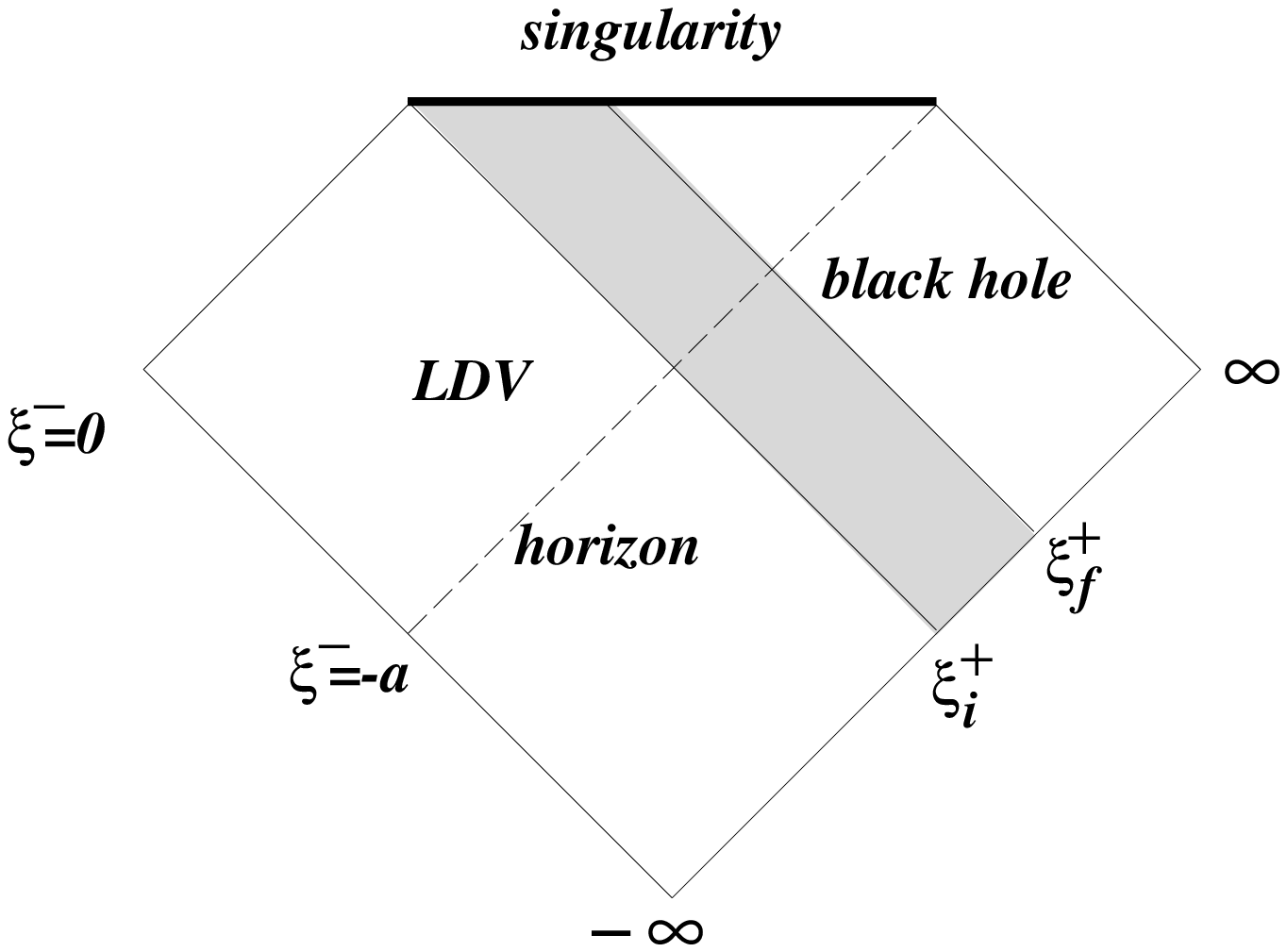}
\end{center}
\par\qquad 
\begin{minipage}{13cm}
{\small {\bf Fig.6} \quad Penrose diagram for  a black hole generated
 by a matter flux. (LDV stands for linear dilaton vacuum.) }
\end{minipage}
\parbigskipn
A particularly simple and useful choice of the flux is 
 the so called \lq\lq shock wave" configuration described by 
the $\delta$-function 
\eqabegin
 F(v) &=& a\delta (\xip -\xipz) \comma \\
M&=& \xipf a-b = a\xipz\period
\eqaend
Clearly the transient region is compressed into a line 
and many simplifications occur. \parsmallskip
In the space-time with a black hole generated by a matter flux, 
there are two distinct asymptotically flat regions and hence 
 two corresponding (asymptotically) flat coordinates. In the  
 \lq\lq in" region,  where we have the linear dilaton vacuum,  
we introduce  $(\zetap, \zetam)$ by 
\eqabegin
 \xip &=& \e^\zetap \qquad \xim =-\e^{-\zetam} \comma \\
 ds^2 &=& -d\zetap\, d\zetam \comma 
\eqaend
 while for the \lq\lq out" region along the future right null infinity 
 $I^+_R$, the proper coordinates $(\sigmap, \sigmam)$ are 
\eqabegin
\xip &=& \e^\sigmap \comma \qquad \xim  = -\e^{-\sigmam} -a \comma \\
ds^2 &=& -{d\sigmap  d\sigmam \over 1+M\e^{-(\sigmap -\sigmam)}} \period
\eqaend
These two coordinate systems are related by the conformal 
 transformation 
\eqabegin
 \bracetwo{ \zetap = \sigmap }{ \zetam
 = -\ln\left( \e^{-\sigmam}+a\right)\period }  
\eqaend
This will be important in discussing the Hawking radiation. 
\parsmallskip
It is instructive to express the parameter $M$ in terms of the 
 energy-momentum tensor as seen by these asymptotic observers. 
 Using the classical transformation property of $T^f_{++}$ and 
 integration by parts, the quantities $a$ and $b$ take the form
\eqabegin
 a &=& \int d\xip T^f_{++}(\xip) =
 \int d\sigmap \e^{-2\sigmap}T^f_{++}(\sigmap)\comma \\
 b&=& \int du \e^u \int^u d\sigmap \e^{-\sigmap}
T^f_{++}(\sigmap) \nn\\
 &=& \e^{\sigmapf} a -\int^{\sigmapf}_{\sigmapi} 
T^f_{++}(\sigmap) \period
\eqaend
Therefore $M$ is expressed as 
\eqabegin
 M &=& \xipf a -b =\int^{\sigmapf}_{\sigmapi}
 T^f_{++}(\sigmap) \period
\eqaend
We see that $M$ is precisely the total matter energy pumped in 
 and,  since the energy conservation holds in the asymptotically 
 flat region, it appears as the mass of the black hole. 
%
\subsection{ Large $N$ Analysis of Hawking Radiation}
Having seen that classically a black hole is generated by a matter flux, 
 we now start incorporating quantum effects upon this configuration.
\parsmallskip
The first question is whether the famous Hawking radiation exists 
 in this model. 
One way of analyzing this problem is to make use of the remarkable 
relation between the trace anomaly and Hawking radiation valid in 
 two dimensions first spelled out by Christensen and Fulling\cite{ChFul}. 
In classically conformally invariant theory in two dimensions, the 
 expectation value of the trace $<T_\mu^\mu>$ can only be 
 proportional to $R$ since no other local scalar of dimension two is 
 available. Indeed by  direct calculation one can show that 
 for models with $N$ massless bosonic fields, one has 
\eqabegin
 < T_\mu^\mu > &=& {N \over 24} R \period
\eqaend
In the light-cone coordinates, $< T_\mu^\mu >= 2g^{+-}<T_{+-}> 
= -4 \e^{-2\rho}<T_{+-} >$ and $ R = 8 \e^{-2\rho}\delplus\delminus\rho 
$ so that the above relation is equivalent to 
\eqabegin
 < T_{+-} > &=& -{N\over 12} \delplus\delminus \rho\period
\eqaend
To show its relevance to the Hawking radiation, 
 Christensen and Fulling first solve $\nabla_\mu T^\mu_\nu =0$
 in the reduced two-dimensional Schwarzschild metric 
\eqabegin
 ds^2 &=& -\left( 1-{2M\over r}\right) dt^2 +\left( 1-{2M\over r}
\right)^{-1} dr^2 \comma 
\eqaend
assuming $T_\mu^\nu $ is $t$ independent.  The resultant $T_\mu^\nu$ 
 has, in general, non-zero trace. Then they compare it with the 
 general form of $T_\mu^\nu $ for massless radiation and find that 
 the anomaly is precisely due to such radiation. \parsmallskip
Let us apply this procedure for the CGHS model\cite{CGHS}.
 The covariant conservation equation reads 
\eqabegin
 0 &=& \nabla^\mu T_{\mu\nu} = g^{\mu\alpha}\nabla_\alpha
 T_{\mu\nu} \nn\\
&=& g^{\mu\alpha} \left( \del_\alpha T_{\mu\nu} - \Gamma^\beta
 _{\alpha \mu} T_{\beta\nu} -\Gamma^\beta_{\alpha\nu}T_{\mu\beta}
\right)  \period
\eqaend
Set $\nu=+$ and evaluate it in conformal gauge with  $<T_{+-}>
 =-(N/12)\delplus\delminus\rho$. We then get a differential equation
 for $<T_{++}> $ ( the one for $T_{--}$ is entirely similar )
\eqabegin
 \delminus <T_{++} >&=& 2\delplus \rho <T_{+-}>
 -\delplus <T_{-+}> \nn\\
&=& -{N\over 12}\delminus\left(
(\delplus\rho)^2 -\delplus^2\rho \right) \period
\eqaend
It is immediately integrated to yield 
\eqabegin
 T_{\pm\pm} &=& -{N\over 12}\left( (\del_\pm\rho)^2
 - \del_\pm^2\rho + t_\pm\right) \comma 
\eqaend
where $t_\pm$ are functions which should be determined from 
 appropriate boundary conditions in the asymptotically flat region. 
A natural boundary condition is that $<T_{\pm\pm}>=0$ in the linear dilaton 
region. Using the form $\e^{-2\rho} = 1+a\e^{\sigmam}$, this yields
\eqabegin
 & & \bracetwo{t_+ =0\comma }{ t_-=
-\left((\delminus \rho)^2-\delminus^2\rho\right)
 =- {1\over 4} \left( 1-
 {1\over \left(1+a\e^{-\sigmam}\right)^2}\right)\period } 
\eqaend
With these $t_\pm$ it is easy to check that at the right null infinity
 $I^+_R$ (\ie as $\sigmap \rightarrow \infty$)
  the components $<T_{++}>$ and $<T_{+-} >$ vanish, while the left-going
 flux is given by 
\eqabegin 
<T_{--}>&\stackrel{\large\sigmap\rightarrow\infty}{\longrightarrow}& 
 {N\over 48} \left( 1-
 {1\over \left(1+a\e^{-\sigmam}\right)^2}\right) \period
\eqaend
This can be interpreted as the Hawking radiation\cite{CGHS}. \parsmallskip
The derivation just presented is elegant but somewhat obscure since 
 the meaning of the expectation value has not been clearly defined. 
A more transparent calculation along the line of the original derivation 
 of Hawking is instructive\cite{GidNel}. The strategy is 
 to first define the out-vacuum $| 0\rangle_{out}$ as seen from the 
observer at $I_R^+$ to be the state in which ${}_{out}\langle 0 | T 
| 0\rangle_{out} =0$ holds, where $T$ stands for $T_{--}$. 
This implicitly determines the normal ordering of $T$. 
Actually, however, what is realized is the in-vacuum $| 0\rangle_{in}$ 
 which does not contain any quanta defined in the linear dilaton 
 region. Since $| 0\rangle_{in}$ and $| 0\rangle_{out}$ are related 
 by a Bogoliubov transformation, what the out-observer sees, namely 
 ${}_{in}\langle 0 | T |0\rangle_{in}$, can be non-trivial. \parsmallskip
The actual calculation can be set up as follows. Let us denote by 
 $x(=\sigmam)$ and $y(=\zetam)$ the coordinate for out and in regions
 respectively. The mode expansion for  the matter field $f(y)$ 
 and the definition of the in-vacuum are given by 
\eqabegin
f(y) &=& \int_0^\infty {d\omega \over \sqrt{2\omega}} \left(
 a_\omega \e^{-i\omega y} + a^\dagger \e^{i\omega y}\right) \comma \\
\left[ a_\omega , a^\dagger_{\omega'} \right] &=& \delta(\omega
 -\omega') \comma \\
a_\omega | 0\rangle_{in}&=& 0 \period
\eqaend
Now we want to evaluate the expectation value ${}_{in}\langle 0 
| T |0\rangle_{in}$ as seen in the out region. Rather than performing the 
 normal ordering explicitly, it is more convenient to regularize 
 the operator product appearing in $T$ by the point-splitting method 
and later subtract out 
 ${}_{out}\langle 0 | T |0\rangle_{out}$. Thus we make a conformal 
 tranformation $y=y(x)$ and define $y_\delta \equiv  y(x+\delta)$
 with $\delta$ infinitesimal. 
Then $T$ in the out-region is given by 
\eqabegin
T(x) &=& \half \delx f(y(x)) \delx f(y(x+\delta)) \period
\eqaend
By a simple calculation, the expectation value in question is computed
 as 
\eqabegin
 {}_{in}\langle 0 | T(x) |0\rangle_{in} &=& \half  {}_{in}\langle 0 |
\delx f(y) \delx f(y_\delta) |0\rangle_{in} \nn\\
 &=& {1\over 4} y' y'_\delta \int_0^\infty d\omega \omega 
 \e^{i\omega (y_\delta -y +i\epsilon)} \nn\\
 &=& -{1\over 4} {y'y'_\delta  \over (y_\delta -y)^2} \comma 
\eqaend
where the prime means  $\del/\del x$. We now expand this expression 
 in powers of $\delta$. If we define 
\eqabegin
\rone &\equiv & {y'' \over y'}\comma \qquad \rtwo \equiv 
 {y''' \over y'} \comma 
\eqaend
we get the expansion
\eqabegin
 y'_\delta &=& y'\left( 1+\delta \rone + \half \delta^2 \rtwo + \cdots 
\right) \comma \\
 (y_\delta -y)^{-2} &\simeq & \delta^{-2} y'^{-2}\left( 1-\delta\rone
 +\delta^2\left({3\over 4} \rone^2 -{1\over 3} \rtwo\right) \right)
\comma \\
  -{1\over 4} {y'y'_\delta  \over (y_\delta -y)^2} 
 &\simeq & -{1\over 4\delta^2} -{1\over 24} \left( \rtwo -{3\over 2}
 \rone^2 \right)
\nn \\
&\simeq & -{1\over 4\delta^2} -{1\over 24} \left\{ y(x), x\right\} 
\comma 
\eqaend
where 
\eqabegin
 \left\{ y(x), x\right\} 
 &=& \del^2_x \ln y'-\half(\del_x \ln y')^2 
\eqaend
is the Schwarzian derivative. 
One should note that this calculation is identical to that of 
 the conformal anomaly for the Virasoro operator. As for 
 ${}_{out}\langle 0 | T | 0\rangle_{out} $, the Schwarzian derivative term
 is absent and we simply get 
\eqabegin
 {}_{out}\langle 0 | T | 0\rangle_{out}  &=&  -{1\over 4\delta^2}\period
\eqaend
As stated before the normal ordering with respect to 
  $| 0\rangle_{out} $ is effected by subtracting this divergence.
 Thus for $N$ matter fields, we finally get 
\eqabegin
  {}_{in}\langle 0 | T^{f,reg}_{--}(x) |0\rangle_{in} &=& 
-{N\over 24} \left\{ y(x), x\right\} \nn\\
&=& -{N\over 24} \left(\left(\ln \zetam'\right)'' -\half 
 \left(\ln \zetam'\right)^2\right) \nn\\
&=&  {N\over 48} \left( 1-
 {1\over \left(1+a\e^{-\sigmam}\right)^2}\right) \comma 
\eqaend
which gives exactly the same Hawking radiation as obtained previously. 
\subsection{ Effect of Back Reaction and Difficulties }
Although it is gratifying that the model is capable of describing 
  Hawking radiation, the expression we have got cannot be the 
 correct one: It gives a divergent answer when integrated over 
 the future right null infinity. The cause of this problem is 
 obvious. As the black hole emits radiation its mass must diminish,  
 but this effect has not been taken into account. A possible way 
 to tackle this back reaction problem is to incorporate some 
 quantum effects in the form of an effective action and re-solve the 
 equations of motion and energy-momentum constraints. \parsmallskip
The simplest of such approximation schemes is to consider the limit of 
 large 
 $N$ while keeping  $N\e^{2\phi}$ finite. This has two merits: (i) Large 
 $N$ corresponds to small $\e^{2\phi}$, hence weak coupling, and 
 one-loop contribution should dominate.  (ii) The only one-loop 
 contribution of order $N$ is the anomaly due to the matter fields, and 
 one can avoid the quantization of the dilaton-gravity sector which
 is rather difficult. Thus, with the anomaly term of $\calO(N)$, 
the effective action becomes 
\eqabegin
S &=&{4\over \gamma^2}\int d^2\xi \Biggl\{ \e^{-2\phi}
\left[ 2\delplus\delminus \rho
 -4\delplus\phi\delminus\phi + \lambda^2\e^{2\rho}\right] \nn\\
 & & \quad -{N\over 12} \delplus\rho\delminus\rho 
+ \half\sum \delplus f_i \delminus f_i \Biggr\} \period
\eqaend
It turns out that the equations of motion following from this action 
can no longer be solved in closed form.  Moreover, the system becomes 
 singular in a certain region. A way to see this is to look at the 
 kinetic matrix (\lq\lq target space metric") and its 
 determinant:
\eqabegin
 K &=& -\matrixii{4\e^{-2\phi}}{2\e^{-2\phi}}{2
\e^{-2\phi}}{{N\over 12}}  \comma \\
{\rm det}\, K &=& 4\e^{-2\phi} \left( {N\over 12} -\e^{-2\phi}\right) 
 \period
\eqaend
Evidently the determinant vanishes for $ \e^{-2\phi} = {N\over 12} $, 
 which occurs within the weak coupling region. \parsmallskip
Dispite these difficulities, various attempts have been made to extract
 some physical consequences from this model\cite{RS1}-\nocite{BD}
\nocite{STh}\nocite{RST}\nocite{HW2}\nocite{BG}\nocite{St}\nocite{Mi1}
\nocite{LO}\nocite{HW3}
\nocite{PS}\nocite{STU}\cite{HW4}. 
For instance, it has been 
 shown that an apparent horizen (\ie boundary of 
 trapped points) forms outside the event horizon and approaches the 
latter as time goes on.  This can be regarded as an indication of the 
 reduction of the mass of the black hole. Another information is 
 that, at least in this model, the classical black hole singularity is
  not resolved by the large $N$ quantum  effects. The most 
 discouraging feature of this approximation is that it is bound to 
 fail in the most interesting regime near the end point of the 
 evaporation. In that region $M$ becomes very small compared with 
$\lambda$ and hence the coupling constant $g^2_s \sim \lambda /M$ 
 becomes large, as was already mentioned in (\ref{gsnearhor}). 
\section{ Solvable Models of Dilaton Gravity}
\sectionnumbering
We have seen that although the CGHS model appears to have many 
attractive features it is hard to go beyond the large $N$ (weak
 coupling) approximation. In such an approximation 
 dilaton-gravity degrees of freedom are not quantized and furthermore
 one cannot satisfactorily discuss what happens in the the strong 
coupling region near the horizon, especially for small $M$. 
 Thus, it is natural to seek some variants of the CGHS model which 
 are fully quantizable. In this section, we will discuss a strategy for 
finding them and describe an application of  such an idea.    
\subsection{ Freedom in the Choice of the Model }
In searching for a good candidate, one must bear in mind that there is 
 an immense freedom in the possible form of the action for dilaton 
gravity models in two dimensions: Since the dilaton field is 
 dimensionless it is easy to construct an infinite number of 
 models which are power-counting renormalizable. Thus, including 
 the counter terms, the form of the action is highly ambiguous. 
 Also as long as the general covariance is respected, one can choose 
 a variety of functional measures.  With this in mind, a large 
 class of models can be represented as a non-linear 
 sigma model, familiar in string theory,  of the form \cite{Al}, \cite{GS} 
\eqabegin
 S &=& -{1\over \gamma^2} \int d^2\xi \sqrt{-\ghat}\, 
 \Biggl( \half G_{\mu\nu}(X)\ghat^{\alpha\beta}\del_\alpha
 X^\mu \del_\beta X^\nu  +\Phi(X)\hat{R} + T(X) \Biggr) \comma 
\eqaend
where $X^\lambda$ are coordinates in the  $\rho-\phi$ space and,
 in the string theory language,  $G_{\mu\nu}(X), \Phi(X)$ and $T(X)$ 
are, respectively, the target space metric, the (target space) dilaton 
 and the tachyon fields. 
\parsmallskip
Of course not every sigma model can be interpreted as a 
 model of gravity in two dimensions: It must possess two dimensional 
 general covariance. In the conformal gauge 
$g_{\mu\nu} = \e^{2\rho}\ghat_{\mu\nu}$, this requirement means
 that the action should be independent of the separation of the 
conformal factor and the background. More specifically, the action 
 should be invariant under the transformation 
\eqabegin
\ghat &\longrightarrow & \e^{2\delta\omega} \ghat \comma \\
 \rho &\longrightarrow & \rho -\delta\omega \comma \\
  \phi & \longrightarrow & \phi  \period
\eqaend
In the string theory context, the consequence of this requirement 
 is well-known: The $\beta$ functions for the (coupling) fields 
 $G_{\mu\nu}$, $\Phi$ and $T$ must vanish. Up to one loop, these 
 equations take the form
\eqabegin
 \beta^G_{\mu\nu} &=& 2\nabla_\mu\nabla_\nu \Phi + \calR_{\mu\nu}
-\nabla_\mu T \nabla_\nu T+\cdots =0\comma \\
 \beta^\Phi &=&  4 ( \nabla\Phi)^2 -4\nabla^2\Phi -\calR + {N-24 \over 3}
\nn \\ 
& &\quad + (\nabla T)^2-2T^2 +\cdots=0 \comma \\
 \beta^T &=& 4\nabla_\mu\Phi \nabla^\mu T -4T
-2\nabla^2 T +\cdots=0 \comma 
\eqaend
where the covariant derivatives and the curvatures are those in the 
 target space. 
\subsection{ A Solvable Models  for $N\ne 24$ }
Let us give an example of solvable models of this type proposed 
 by de Alwis\cite{Al} and independently by Bilal and Callan\cite{BC}.
 As the sigma model aspect is clearer, we follow \cite{Al}.  First 
 assume that for weak coupling $\e^{2\phi} << 1$ the model reduces to
 that of CGHS with an addition of an anomaly term, namely to 
\eqabegin
 S &=& {1\over \gamma} \int d^2\xi \sqrt{-\hat{g}}\, 
 \Biggl\{ \e^{-2\phi} \left[ \hat{R} + 4(\hat{\nabla}\phi)^2 -
4\hat{\nabla}
\phi\cdot \hat{\nabla}\rho +4\lambda^2\e^{2\rho} \right]  \nn\\
& & + \kappa \left( (\hat{\nabla}\rho)^2 + \hat{R}\rho \right)
\Biggr\} \period
\eqaend
In terms of the sigma model functions, this means 
\eqabegin
 G_{\phi\phi} &=& -8\e^{-2\phi}\comma\quad G_{\phi\rho}= 4\e^{-2\phi}
\comma \quad G_{\rho\rho} = -2\kappa \comma \nn\\
\Phi &=& -\e^{-2\phi} - \kappa\rho\comma \quad T = -4\lambda^2
 \e^{2(\rho-\phi)} \period 
\eqaend
Now introduce functions $h(\phi)$, $\bar{h}(\phi)$ and $\bar{\bar{h}}
(\phi)$ of order $\e^{2\phi}$ which express deviations of $G_{\mu\nu}$
 from the CGHS form and write the target space line element as 
\eqabegin
 ds^2 &=& -8\e^{-2\phi} \left[ 1+h(\phi)\right] d\phi^2  \nn\\
&&  \quad + 8\e^{-2\phi} \left[ 1+\bar{h}(\phi)\right] d\rho d\phi \nn\\
&& \qquad -2\kappa ( 1+\bar{\bar{h}}(\phi)) d\rho^2 \period
\eqaend
It can be shown that for $\bar{\bar{h}}=0$ the target space curvature 
$\calR$ vanishes.  Restricting to that case, there must exist a target 
 space coordinate system in which the metric becomes manifestly flat. 
It turns out that the following transformation $(\rho, \phi) 
 \rightarrow (x,y)$ does the job:
\eqabegin
 y &\equiv & \rho +{1\over \kappa}\e^{-2\phi} 
 -{2\over \kappa}\int d\phi \e^{-2\phi}\bar{h}(\phi) \comma \\
 x &\equiv& \int d\phi \e^{-2\phi}\left( (1+\bar{h})^2
 -\kappa\e^{2\phi}(1+h)\right)^{1/2} \comma \\
 ds^2 &=& {8 \over \kappa} dx^2 -2\kappa dy^2 \period
\eqaend
Next, de Alwis analyzes the $\beta$ function equations with 
 small $T$ approximation. The solution so obtained can be summarized 
 in the following form of the action (including the massless 
 matter fields $f_i$)
\eqabegin
S &=&{1\over \gamma^2} \int d^2\xi \Biggl\{
\mp\left(\delplus X \delminus X-\delplus Y\delminus Y\right) 
+\sum_i \delplus f_i \delminus f_i \nn\\
&&\quad + 2\lambda^2\e^{\mp
\sqrt{2/|\kappa|}\, (X\mp Y)} \Biggr\} \comma 
\eqaend
where
\eqabegin
\kappa &=& {N-24 \over 6} \comma \\
 X&\equiv & 2\sqrt{2/|\kappa|}\ x\comma \qquad Y \equiv
 \sqrt{2|\kappa|}\ y \period
\eqaend
It can be shown that by a further transformation this action 
 can be mapped on to a conformally invariant system described in terms
 of free fields. Such a system is obviously classically solvable (with the 
 inclusion of the anomaly term).  Futhermore it is apparently solvable even 
 fully quantum mechanically. However, there is a 
 serious problem: Since the transformation $(\rho,\phi) 
\longleftrightarrow (X,Y)$ shown above is highly non-linear and non-local, 
 the relation between the original and the transformed operators 
 as well as the functional measure taken are quite obscure. For 
 this reason, essentially classical analysis has been performed for 
these models. \parsmallskip
{\it Note added}: After this talk was delivered, we have found that 
 a large class of solvable models (including those discussed above 
 and the one which will be treated in detail in the next two sections) 
can be characterized as non-linear sigma models {\it with a special 
symmetry}. (This symmetry  is an extensive generalization
 of the one discussed in \cite{RST}.)  
With that observation,  the discussions presented in this 
section can be made much more transparent and general. Interested
 reader should be referred to \cite{KS3}. 
\section{ An Exactly Quantizable Model of CGHS Type for 
$N=24$} 
\sectionnumbering
As we have seen, there are enormous possibilities for models 
 of dilaton gravity in two dimensions, even within  solvable models.
In this section, we shall discuss one particular model, for 
 which exact quantization can be carried out and all the  physical states 
 can be constructed \cite{HKS}. 
 An attempt to extract the averaged geometry they 
 describe, including that of a black hole, will be presented in 
 the next section. 
\subsection{ Choice of the Model }
In order to define the quantum model we deal with,  first recall
 the classical CGHS model in the \lq\lq canonical" metric 
$\gtil_{\mu\nu}$ discussed before:
\eqabegin
S &=& \ff \int d^2\xi \sqrt{-\gtil}\, \left( \Phi \Rt + 4\lambda^2
-\half \sum(\tilde{\nabla} f_i)^2
\right) \comma \\
 \Phi &\equiv & \e^{-2\phi} \comma \qquad 
 g_{\mu\nu} = \Pinv \gtil_{\mu\nu} \period
\eqaend
As was remarked at that time, the kinetic term for $\Phi$ is absent 
 in this expression.  We now  reinstate it by a 
 conformal transformation of the form
\eqabegin
\gtil_{\mu\nu} &=& \e^\Phi  h_{\mu\nu} \comma \\
\sqrt{-\gtil} &=& \e^\Phi \sqrt{-h} \comma \\
\tilde{R} &=& \e^{-\Phi} \left( R_h -h_{\mu\nu}\nabla_\mu \nabla_\nu
 \Phi \right) \period
\eqaend
This results in a Liouville like action, first written down by Russo and 
Tseytlin \cite{RT0}: 
\eqabegin
S &=& \ff \int d^2\xi \sqrt{-h}\, \Biggl((\nabla \Phi)^2
 + 4\lambda^2\e^\Phi + \Phi R_h 
  -\half \sum(\nabla f_i)^2 \Biggr) \period
\eqaend
To quantize this model, we must specify the functional measure. 
One possible choice is the measure induced by 
 the functional norms of the form \cite{HT}
\eqabegin
 \parallel \delta h \parallel^2 &=& \int d^2x \sqrt{-h}\, 
   h^{\alpha\beta}h^{\gamma\delta}\delta h_{\alpha\gamma}h_{\beta\delta}
 \comma \nn\\
 \parallel \delta\Phi \parallel^2 &=& \int d^2x \sqrt{-h}\,
 \delta\Phi \delta\Phi \comma \\
 \parallel \delta f_i \parallel^2 &=& \int d^2x \sqrt{-h}\,
 \delta f_i \delta f_i  \qquad (i=1,\ldots, N) \comma \nn
\eqaend
which are natural with respect to the metric $h_{\mu\nu}$. 
Now adopt the conformal gauge 
\eqabegin
 h_{\alpha\beta} &= &\e^{2\rho_h}\, \hat{g}_{\alpha\beta}\comma 
\eqaend
and perform the famous  David-Distler-Kawai analysis \cite{DDK} to
 change to the standard translationally invariant measure. The 
 result is the effective action
\eqabegin
S &=& \ff\int d^2\xi \sqrt{-\ghat}\, \Biggl[\, 
 \nablah\Phi\hat{\cdot}\nablah\Phi +2\nablah\Phi
 \hat{\cdot}\nablah \rho_h +\hat{R}\Phi\nn\\
 & &\quad  +4\lambda^2 \e^{\Phi + 2\rho_h}
 + \gamma^2 {N-24 \over 24\pi} \left( \nablah\rho_h
 \hat{\cdot}\nablah\rho_h + \hat{R}\rho_h\right) \nn\\
 & &\quad\quad
  -\half\nablah\vec{f} \hat{\cdot} \nablah\vec{f}
\, \Biggr] + S^{gh}(\ghat, b,c) \comma 
\eqaend
where hatted quantities are computed with the reference metric 
$\ghat_{\mu\nu}$ and $S^{gh}$ is the usual ghost action. 
Although everything is clear and straightforward, the procedure above 
is not without problems. One, which is frequently encountered in other 
dilaton gravity models as well, is that  the positivity of $\Phi$ is 
not properly respected. Another problem is that it is not clear in which 
conformal frame we should interpret the model physically.  If we 
 choose to interpret the physics in the original frame, namely 
 with respect to $g_{\mu\nu}$, the classical black hole singularity 
 occurs at $\Phi=0$ and flucutations in the region of negative $\Phi$, 
 the first problem,  might not be so relevant.  In any case, as we do 
not have satisfactory answers to these problems, we shall proceed 
 with the model and watch if anything pathological will happen. 
Furthermore, we will focus on the special case with $N=24$ 
 in which the term due to the anomaly drops out and the model 
 simplifies considerably. Essentially the 
 same model was analyzed from a different point of view in 
\cite{VV}. 
\parsmallskip
To perform a rigorous quantization, we will put the system in a 
 \lq\lq box" (\ie an interval) of size $L$ and impose 
 the periodic boundary conditions. This is best implimented by 
introducing the dimensionless variables
\eqabegin
 x^\alpha &= & (t,\sigma) = \xi^\alpha / L\comma\qquad \mu=\lambda L 
\comma 
\eqaend
and demand that all the fields appearing in the action be 
$2\pi$-periodic in $\sigma$. \parsmallskip
As far as the classical analysis is concerned, it is identical
 to that for the CGHS model.  For convenience, let us display  
 the relevant results obtained before (see Eqs.(\ref{gmunu}) $\sim$ 
 (\ref{chiitoT})), with the notational replacements
 $\rho\longrightarrow \rho_g$, $\rho_h \longrightarrow\rho$. 
\eqabegin
g^{\mu\nu} &=& \e^{-2\rho_g}\eta^{\mu\nu} =\Phi\e^{-\psi}\eta^{\mu\nu}
=\Phi \e^{-(\Phi+2\rho)}\eta^{\mu\nu} \comma \\
\psi &=&\psip(\xplus)+\psim(\xminus) =\mbox{ free field} \comma \\ 
\rho &=& \half (\psi -\Phi) \comma\\
\Phi &=& \e^{-2\phi} = -\left(\chi +AB\right) \comma\\
\chi &=& \chi_+(\xip) + \chi_-(\xim) =\mbox{ free field}\comma\\
\delplus A &=& \mu\e^{\psip(\xplus)}\comma \qquad 
\delminus B = \mu\e^{\psim(\xminus)}\comma\\
\gt^2 T_{\pm\pm} &=& \del_\pm \chi \del_\pm \psi -\del_\pm^2\chi + 
\half \sum (\del_\pm f_i)^2 \comma \label{emtensor} \\
\gt &\equiv & {\gamma \over \sqrt{4\pi}} \period
\eqaend
\parsmallskip
Previously, the solutions for $A(\xplus)$ and $B(\xminus)$ were 
 considered only in the $\psi=0$ gauge. As we wish to utilize the 
 full conformal invariance, we do not want to impose that 
 restriction here.  Moreover, we need to solve for these functions 
 with proper boundary conditions. In fact this problem is exactly the 
same as the one which occurs in the operator treatment of the Liouville 
theory \cite{OW}, \cite{KN}.  
To find the boundary conditions for $A$ and $B$, 
 we write out the mode expansion for $\psi$. For the left-moving 
 part, it can be written as 
\eqabegin
 \psi^+ &=& \gt\left\{ {\qplus \over 2} + \pplus x^+
  + i\sum_{n\ne 0}{\alpha^+ _n\over n} \e^{-inx^+}\right\}
  \period
\eqaend
 and similarly for $\psi^-$. 
Under $\sigma \rightarrow \sigma + 2\pi$, 
$\psi^\pm$ undergoes a shift which depends on the  {\it  zero mode}
$p^+$. This means that although the product $AB$ is periodic $A$ and 
 $B$ separately are not  and they undergo the change
\eqabegin
 A(\xplus +2\pi) &=& \alpha A(\xplus)  \comma \\
 B(\xminus -2\pi) &=& {1\over \alpha} B(\xminus) \comma 
\eqaend
where $\alpha = \e^{\gamma \sqrt{\pi}\pplus }$. The solutions satisfying
 these boundary conditions  take the form (suppressing 
 the $t$ dependence)
\eqabegin
 A(\sigma) &=& \mu C(\alpha) \int_0^{2\pi} d\sigma'
    E_\alpha(\sigma-\sigma') \e^{\psi_+(\sigma')}  \comma \\
 B(\sigma) &=& \mu C(\alpha)\int_0^{2\pi} d\sigma''
    E_{1/\alpha}(\sigma-\sigma'') \e^{\psi_-(\sigma'')} \period  
\eqaend
Here  $C(\alpha) = 1/\left(\sqrt{\alpha}-\sqrt{\alpha}^{-1}\right)$
 and the function $E_\alpha(\sigma)$ is a variant of 
a step function, shown in  Fig.7,  given by 
\eqabegin
  E_\alpha(\sigma) &\equiv& 
      \exp\left(\half\ln\alpha\, \epsilon(\sigma)\right)  \period
\eqaend
$\epsilon(\sigma)$ is the usual stair step function with the 
 property $\epsilon(\sigma +2\pi) = 2 +\epsilon(\sigma)$. 
%
\begin{center} 
\epsfxsize=8cm
\quad\epsfbox{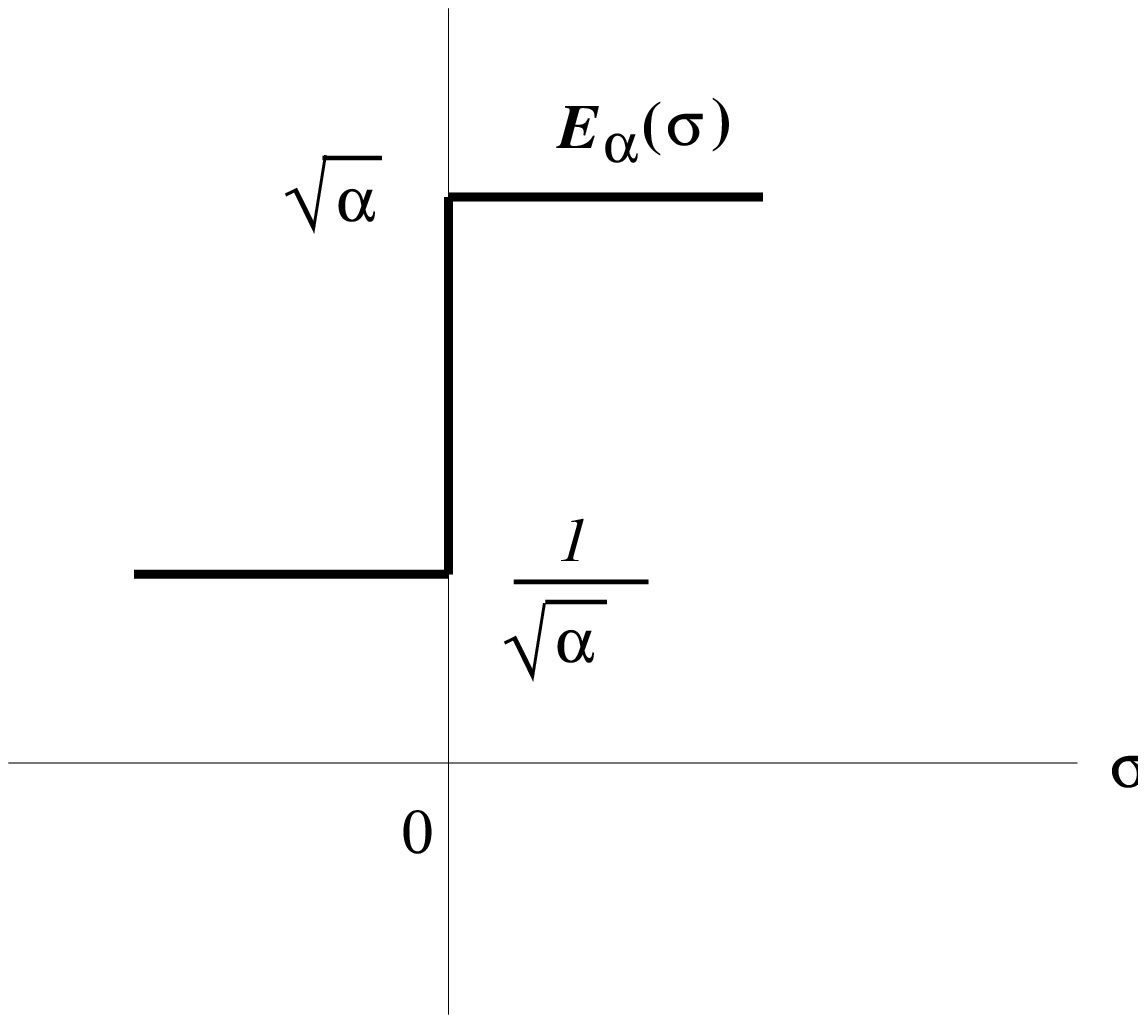}
\end{center}
\begin{center}
{\small {\bf Fig.7} \quad Sketch of the step function $E_\alpha(\sigma)$.}
\end{center}
\parbigskipn
Let us now go back to the energy-momentum tensor $T_{\pm\pm}$ given 
 in (\ref{emtensor}). Introduce $\phi_1,\phi_2$ and $\phi^i_f$ by 
\eqabegin
 \psi &=& {\gt \over \sqrt{2}} (\phi_1 +\phi_2)\comma 
    \qquad 
 \chi ={\gt \over \sqrt{2}} (\phi_1 -\phi_2)
 \comma \\
 f^i &=& \gt \phi^i_f \period
\eqaend
Then $T_{\pm\pm}$ become diagonal:
\eqabegin
  T_{\pm\pm}&=& \half( \del_\pm \phi_1 )^2 
-Q \del_\pm^2 \phi_1 
  -\half( \del_\pm \phi_2 )^2 +Q\del_\pm^2
 \phi_2  +\half (\del_\pm \vec{\phi}_f)^2  \comma 
\eqaend
where the background charge $Q$ is given by
\eqabegin 
Q &=& {\sqrt{2\pi}\over \gamma}\period \label{bgc}
\eqaend
Note $\phi_2$ is of {\it negative metric}. \parsmallskip
It is not difficult to obtain the general solutions of $T_{\pm\pm}=0$ 
 satisfying the proper boundary conditions. To display them, it is 
 convenient to write $A(\xplus)$ and $B(\xminus)$ in the form 
\eqabegin
 A(\xplus) &=& \mu \e^{\gt \pplus \xplus} a(\xplus)\comma \\
 B(\xminus) &=& \mu \e^{\gt\pplus \xminus}b(\xminus)\comma 
\eqaend
where $ a(\xplus)$ and $b(\xminus)$ are arbitrary periodic functions. 
Then the general solutions are spanned by $ a(\xplus), b(\xminus)$
 and $\chi$ and $\psi$, $T^f_{\pm\pm}$ and $g_{\alpha\beta}$
 are expressed  in terms of them. In particular, the original 
metric takes the form 
\eqabegin
 g_{\alpha\beta} &=& {\e^{\psi}\over \chi +AB } \eta_{\alpha\beta}\nn\\
 &=& {\e^{\psi}\over \chi + \mu^2 \e^{2\gt\pplus t}a(\xplus)b(\xminus)}
 \eta_{\alpha\beta}  \period
\eqaend
Let us give an example. A configuration which describes a 
matter-free black hole in the large $L$ limit can be produced by the choice 
\eqabegin
 \chi &=& -c = \mbox{constant}, \qquad \del_\pm \vec{f} = 0 \comma \\
 a(\xplus) &=& \sin\xplus,\qquad b(\xminus) = \sin\xminus \comma 
\eqaend
for which $\psi$ becomes 
\eqabegin
 \psi_\pm &=& \gt\pplus x^\pm + \ln \left(\gt\pplus
 \sin x^\pm + \cos x^\pm \right) \period
\eqaend
Recalling the definitions $x^\alpha = \xi^\alpha /L$ and 
 $\mu =\lambda L$, we indeed get 
\eqabegin
 \lim_{L\rightarrow \infty} g_{\alpha\beta} &=& 
 \lim_{L\rightarrow \infty} { - \e^{\psi}\over
c -\mu^2 \e^{2\gt\pplus t}\sin \xplus
 \sin \xminus}\eta_{\alpha\beta}  \nn\\
  &=& -{1\over c -\lambda^2 \xi^+ \xi^- }\eta_{\alpha\beta}
  \comma 
\eqaend
which describes a simple black hole. A black hole produced by a matter 
 shock  wave can similarly be constructed. 
\subsection{ Quantization of the Model}
Since the original fields are expressed in terms of free fields, 
 we expect that the usual quantization of the latter achieves 
 the full quantization of the model. We can confirm this by showing 
 that the non-linear and non-local transformation $(\Phi, \rho)
 \rightarrow (\psi, \chi)$ is a canonical  transformation 
quantum mechanically as well  as classically. 
\parsmallskip
Let us write the mode expansion for the free bosons $\phi_i,\ (i=1,2)$
 as 
\eqabegin
 \phi_i(\xplus,\xminus) &=& \phi_i^+(\xplus) + \phi_i^-(\xminus)\comma \\
 \phi_i^\pm(x^\pm) &=& {q^i \over 2} + p^i x^\pm
    + i\sum_{n\ne 0} {1\over n} \alpha_n^{(i,\pm)}
    \e^{-inx^\pm}  \comma 
\eqaend
 and assume the standard free field Poisson bracket relations :
\eqabegin
  i\left\{ \alpha_m^{(1,\pm)}, \alpha_n^{(1,\pm)} \right\} &=&
  m\delta_{m+n,0}\comma \qquad \left\{q^1, p^1 \right\} = 1 \comma \\
   i\left\{ \alpha_m^{(2,\pm)}, \alpha_n^{(2,\pm)} \right\}
 &=& - m\delta_{m+n,0}\comma\qquad \left\{q^2, p^2 \right\} = -1
 \comma \\ 
 \mbox{Rest} &=& 0  \period
\eqaend
One can then easily compute the general Poisson brackets between 
 the fields $\psi$ and $\chi$.  The result shows that, as can be guessed 
 from the form of $T_{\pm\pm}$,  $\psi$ and 
 $\chi$ are conjugate to each other and 
 $\left\{ \psi(x), \psi(y)\right\} = \left\{ \chi(x), \chi(y)\right\}
    =0$. Now express the original fields $(\Phi, \rho)$ and their
 conjugate momenta $(\Pi_\Phi, \Pi_\rho) = (2 (\Phidot + \rhodot),
2 \Phidot)$ in terms of $(\psi, \chi)$ and compute the 
 equal time (ET) Poisson brackets among them. Calculations are 
 tedious due to the presence of non-local expressions and zero mode 
 factors, but they lead to the expected result
\eqabegin
  \left\{ \Phi(x),\Pi_\Phi(y) \right\}_{ET}
    &=& \delta(\sigma_x -\sigma_y) \comma \\
\left\{ \rho(x), \Pi_\rho(y) \right\}_{ET} 
    &=& \delta(\sigma_x -\sigma_y) \comma \\ 
 \mbox{Rest} &=& 0 \comma 
\eqaend
showing that the transformation is canonical at least classically. 
\parsmallskip
To quantize the system, we intend to make the usual replacement 
$ \left\{\phi_i, \phi_j\right\}_{ET} \rightarrow 
  {\hbar \over i}\left[\phi_i, \phi_j\right]$. When the transformation
 is non-linear, its canonical nature can be 
 destroyed in this process due to the normal-ordering necessary for the 
composite operators involved. Fortunately for the present model this 
 does not happen: The dangerous composite operator $AB$ consists 
 only of $\psi$ field and its components commute among themselves as
 we remarked earlier. Thus {\it normal ordering is unnecessary} and 
 the proof of canonicity goes through exactly as in the classical case.
 In this respect, this model is far simpler than the 
 bona fide Liouville theory.
\parsmallskip
After the quantization the model continues to enjoy conformal 
 invariance without anomaly. The Virasoro generators $L^{dL}_n$ 
 and $L^f_n$,  for the dilaton-Liouville (dL) and the matter (f)
 sector respectively, are given by 
\eqabegin
 \Ldl_n &=& L_n^1 + L_n^2  \comma  \\
 L_n^1 &=& +\half \sum_m :\alpha^1_{n-m}\alpha^1_m:+iQn\alpha^1_n
\comma \\
 L_n^2 &=& -\half \sum_m :\alpha^2_{n-m}\alpha^2_m:-iQn\alpha^2_n
\comma \\
 \Lf{n} &=&  \half \sum_m :\vecal^f_{n-m}\cdot\vecal^f_m:
\comma 
\eqaend
where $Q$ is as given in (\ref{bgc}). Note the overall negative sign 
 for $L_n^2$ due to the negative metric associated with the oscillators 
$\alpha^2_n$. When we compute the algebra of these
 operators, the $Q$-dependence cancel out and we find the 
 usual form given by 
\eqabegin
 \left[\Ldl_m, \Ldl_n \right] &=& (m-n)\Ldl_{m+n}
  + {2\over 12}(m^3-m)\deltamn  \comma \\
 \left[\Lf{m}, \Lf{n} \right] &=& (m-n)\Lf{m+n}
  + {N\over 12}(m^3-m)\deltamn  \period
\eqaend 
Evidently for $N=24$ the central charges add up to the critical value 
 $c=26$. With respect to the total generator $L^{tot}_n$, the free fields 
 $\psi$ and $\chi$ transform as 
\eqabegin
 \left[L^{tot}_m, \psi(x)\right] &=& \e^{im\xplus} 
 \left( \overi \delplus \psi +  im \right) 
\comma \\
 \left[L^{tot}_m, \chi(x)\right] &=& \e^{im\xplus} 
  \overi \delplus \chi  \period
\eqaend
Thus,  while $\chi$ is a genuine conformal field with dimension $0$, 
 $\psi$ transforms with an additional inhomogeneous term. 
 Because of this contribution, 
$ \mbox{dim}\ \e^{\lambda\psi}  = \lambda$ and in particular the 
operator $\e^\psi$ appearing 
 in the action is a $(1,1)$ operator. As for $\e^{\lambda\chi}$, 
its dimension is $0$ regardless of $\lambda$. 
Another operator of importance is $AB$. By a somewhat tedious 
calculation, one can show that, despite its complexity, it is a 
dimension 0 primary:
\parn
\eqabegin
 \left[ L_n, AB(x)\right] &=& \e^{in\xplus}
 {1\over i} {\del \over \del\xplus} (AB(x))
  \period
\eqaend
This is desirable since we want $\Phi = -(\chi + AB)$ to have 
 a definite transformation property. 
\subsection{ BRST Analysis of Physical States }
Having quantized the model exactly, the next task is 
 to construct the physical states\cite{HKS}. 
 As the reader must have noticed, the 
mathematical structure of the model is identical to that of a 
bosonic string theory in 26 dimensions {\it  with a special background 
charge.} Thus techniques developed for string theory can be 
 readily transcribed to this case with minor modifications. 
Specifically, we shall apply  the BRST method which was quite 
 successful in analyzing the physical states of string models 
 with central charge $c\le 1$\cite{LZ}\cite{BMP}. (Analysis similar 
to the one below was also performed in 
\cite{Bilal}\cite{Sakaietal}.) \parsmallskip
We shall explicitly deal only with the left-moving sector 
 and use chiral fields $\phi_1(\xplus)$ and  $\phi_2(\xplus)$
 previously introduced. The mode expansion and the basic commutation 
relations are given by 
\eqabegin
 \phi_i (\xplus) &=& \half q_i + p_i\xplus + i\sum_{n\ne 0}
 {1\over n} \alpha_n^i \e^{-in\xplus} \comma \\
 \delplus \phi_i &=& \sum_{n\in {\bf Z}} \alpha_n^i \e^{-in\xplus} 
 \qquad ( \alpha^i_0 \equiv p^i )  \comma \\
 \left[ q_1, p_1 \right] &=& i\hbar, \qquad 
 \left[ q_2, p_2 \right] = -i\hbar \comma \\
 \left[ q_1, p_2 \right] &=& \left[ q_2, p_1 \right]=0 \comma \\
 \left[ \alpha_m^1, \alpha_n^1 \right] &=& 
 - \left[ \alpha_m^2, \alpha_n^2 \right] = m\hbar\deltamn , \qquad 
 \left[ \alpha_m^1, \alpha_n^2 \right] = 0  \period
\eqaend
In the standard fashion we can construct the BRST operator $d$ as 
\eqabegin
 && d = \sum c_{-n} (\Ldl_n+\Lf{n}) -\half 
\sum :(m-n)c_{-m}c_{-n}b_{m+n} : \comma 
\eqaend
where the normal ordering for the ghost and the anti-ghost oscillators,
 $c_n$ and $b_n$, is with respect to the so-called 
\lq\lq down vacuum" $\downvacket$, which is related to the 
$SL_2$-invariant vacuum by $ \downvacket = c_1 \mid 0 >_{inv}$. 
\parsmallskip
In the following, our emphasis will be on the general strategy
 for finding physical states and therefore we often omit  technical 
 details. In the BRST formalism, a physical state is characterized 
 as a state 
 which is annihilated by $d$ (BRST-closed) yet cannot be written 
 as  $d\ket{\Lambda}$ for some state $\ket{\Lambda}$ 
(BRST-non-exact).  Thus, a physical state $\ket{\Psi}$ can always be 
 written in the form
\eqabegin
 \ket{\Psi} &=& \ket{\Psizero} + d\ket{\Lambda} \comma 
\eqaend
where $\ket{\Psizero}$, which cannot be of the form $d\ket{\ast}$, 
 will be referred to as the non-trivial part. 
Mathematically, states 
 satisfying these conditions are said to form $d$-cohomology classes
 or {\it absolute cohomology} classes. 
Since $d$ is a rather complicated operator it is hard to find these 
 states directly. The basic idea is to reduce the problem to 
 that of finding simpler cohomologies embedded in the original 
 and with that information reconstruct the absolute 
 cohomology. \parsmallskip
The first step is to decompose the operator $d$ with respect to the 
 ghost zero modes $c_0$ and $b_0$:
\eqabegin
d &=& c_0 L^{tot}_0 -Mb_0 + \dhat \comma 
\eqaend
where $L^{tot}_0$ is the total Virasoro generator at level $0$, 
 $M$ is a complicated operator the structure of which does not concern
 us, and $\dhat$  does not contain $c_0$ nor 
$b_0$. From the above decomposition it is easy to see that 
 $L^{tot}_0$ can be written as an anti-commutator 
\eqabegin
L^{tot}_0 &=& \left\{b_0, d\right\} \period \label{eqn:Lzbd}
\eqaend
This is important since it implies that the non-trivial
 part $\ket{\Psizero}$ of a physical state must satisfy 
$L^{tot}_0 \ket{\Psizero} = 0$. The proof is elementary: Suppose 
 the global weight of  $\Psizeroket$  is $h$, \ie, 
$L^{tot}_0\Psizeroket = h\Psizeroket$. We can rewrite this as 
\eqabegin
L^{tot}_0\Psizeroket &=& h\Psizeroket \nn\\
 &=& (db_0 + b_0 d)\Psizeroket = d(b_0\Psizeroket) \period
\eqaend
Hence if $h\ne 0$ we can divide by $h$ and find that 
$\Psizeroket= d\left({1\over h}b_0  \Psizeroket \right)$, showing 
 that it is not a physical state. As this contradicts the assumption, 
 we must have $h=0$. \parsmallskip
With this in mind, we can now define a subspace of states called 
 the space of {\it relative cohomology} ${\cal F}_0$ as follows:
\eqabegin
{\cal F}_0 &\equiv & \left\{ \ket{\psi}\, \mid\, 
L^{tot}_0\ket{\psi}=0, 
\ b_0\ket{\psi} =0 \right\}\period
\eqaend
Since $b_0$ as well as $L^{tot}_0$ vanishes on this space, we have 
$d=\dhat$ and hence $\dhat^2=0$ on ${\cal F}_0$.  Thus one can 
 consider a cohomology with respect to $\dhat$. \parsmallskip
Although $\dhat$ is simpler than $d$, it is still quite complicated
 and one would like to reduce the problem further. This is achieved 
 by decomposing $\dhat$ according to the so called  {\it degree} 
 assigned to appropriate operators. For this purpose, introduce the 
 following light-cone type variables:
\eqabegin
 \qpm &=& \overrttwo \left( \qone \pm \qtwo \right)\comma  
 \qquad 
 \ppm = \overrttwo \left( \pone \pm \ptwo \right) \comma \\
\left[ \qpm, \ppm \right] &=& i\hbar \comma \\ 
  \alpm_m &=& \overrttwo \left( \alpha^1_m \pm  \alpha^2_m \right) 
 \qquad
 \left[ \alpm_m, \alpha^\mp_n \right] = m\hbar\deltamn \period
\eqaend
To these  operators we  assign the degrees
\eqabegin
 \mbox{deg}(\alplus_n ) &=& \mbox{deg}(c_n) =1\comma   
 \qquad
 \mbox{deg}(\alminus_n ) = \mbox{deg}(b_n) =-1 \comma \\
 \mbox{deg(Rest)} &=& 0 \period
\eqaend
Then  $\hat{d}$ is decomposed according to the degree as 
\eqabegin
 \dhat &=& \dhat_0 + \dhat_1 +\dhat_2 \comma \label{decompdh} \\
 \dhat_0 &=& \sum_{n\ne 0} P^+(n)c_{-n}\alminus_n \comma  \\
 \dhat_1 &=& \sum_{n.z.m.}:c_{-n}( \alplus_{-m}\alminus_{m+n}
     + \half (m-n)c_{-m}b_{m+n} +\Lf{n}) :\comma  \\
 \dhat_2 &=& \sum_{n\ne 0} P^-(n)c_{-n}\alplus_n \comma 
\eqaend
where $n.z.m.$ stands for summation over the non-zero modes and 
\eqabegin
 P^+(n) &=& \overrttwo \left( \pone +\ptwo +2iQn \right) \comma \\
 P^-(n) &=& p_- \period
\eqaend
Notice that although $\dhat_1$ is still rather involved $\dhat_0$ 
 and $\dhat_2$ have very simple structures. Since they have 
 the lowest and the highest degree in the decomposition
 (\ref{decompdh}), $\dhat^2
 =0$ implies $\dzerohat^2 = \dhat_2^2 = 0$. Furthermore, it is easy 
 to see that they commute (anti-commute) with $L^{tot}_0$ ($b_0$). 
 This allows us to consider $\dzerohat$- or $\dhat_2$- cohomologies 
 consistently on ${\cal F}_0$. (Which of these cohomologies is more
 useful depends on the form of $P^\pm(n)$.) Fortunately 
 these cohomologies are simple enough to be solved explicitly. 
 \parsmallskip
Let us give an example. When $P^+(n) \ne 0$  for all non-zero $n$,
 one can construct an operator 
\eqabegin
\Kplus &\equiv& \sum_{n\ne 0} {1\over \Pplus(n)} \alplus_{-n} b_n 
\period
\eqaend
This operator is kind of an inverse of $\dzerohat$. Its anti-commutator
 with $\dzerohat$ yields
\eqabegin
 \left\{ \dhat_0, \Kplus\right\} &=& \sum_{n\ne 0}:(nc_{-n} b_n 
  + \alplus_{-n} \alminus_n ):+\hbar \nn\\
 &= & \hat{N}^{dLg} \comma 
\eqaend
where $\hat{N}^{dLg}$ is precisely the level-counting operator in
 the dilaton-Liouville-ghost (d-L-g) sector.  This relation, which is 
  quite similar to (\ref{eqn:Lzbd}), implies that a non-trivial 
 state in $\dzerohat$ cohomology must satisfy 
\eqabegin
 \hat{N}^{dLg}\ket{\psi} &=& 0 \comma \label{NdLg}
\eqaend
namely that it should not contain any non-zero mode excitations
 in the $d$-$L$-$g$ sector.  Taking this into account, the 
 $L^{tot}_0\ket{\psi}=0$ condition becomes 
\eqabegin
 L^{tot}_0\ket{\psi} &=& \left( \pplus\pminus + \half \vecpf^2+\Nhatf 
-\hbar\right) \ket{\psi} = 0 \period \label{Lzerocond}
 \eqaend
This expresses nothing but the energy balance between the matter 
 and the zero modes of dilaton and gravitational fields. In
 other words, whenever matter is present it gets gravitationally 
dressed such that the total energy of the system vanishes. 
The problem has now been reduced to enumerating states satisfying 
 two conditions (\ref{NdLg}) and (\ref{Lzerocond}) and  this can easily be 
 done. 
\parsmallskip
Reconstruction of the relative and the absolute cohomologies 
 from the $\dzerohat$- (or $\dhat_2$-) cohomologies is a bit 
 involved.  We omit the details and show only the correspondance:
\eqabegin
&& \mbox{$\dhat_{0,2}$-cohomology}\quad 
\stackrel{1:1}{ \Longleftrightarrow} \quad 
  \mbox{$\dhat$-chomology} \quad \stackrel{1:2}{\Longleftrightarrow}
  \quad \mbox{$d$-cohomology} \period
\eqaend
Just to give you a feeling, let us display the formula which 
 constructs the state $\ket{\psi}$ in the relative cohomology out of 
 a state $\ket{\psi_0}$ in the $\dzerohat$-cohomology:
\eqabegin
 \ket{\psi} &=& \sum_{n=0}^\infty (-1)^n (T^+)^n
 \ket{\psi_0} \comma \label{dhatcohom}\\
 T^+ &\equiv& \hat{N}_{dLg}^{-1} K^+ \dhat_1 \period
 \eqaend
Since both $K^+$ and $\dhat_1$ are rather involved, the expression 
 above appears to be extremetly complicated. However, when one 
 starts writing out simple examples, one almost immediately understands 
 what this formula means. Let $\pdownvac$ be a state of zero energy 
 made out of zero modes $p^\pm$ and $\vec{p}_f$ only. Then, for 
 example, 
\eqabegin
\mbox{For}\qquad \ket{\psi_0} &=& \alpha^i_{-1}\pdownvac 
\comma \\
 \ket{\psi} &=& \Biggl[ \alpha^i_{-1} -{p_f^i\over \Pplus(1)}
   \alpha^+_{-1}\Biggr] \pdownvac \comma \\
\mbox{For}\qquad \ket{\psi_0} &=& \alpha^i_{-2}\pdownvac \nn\\
 \ket{\psi} &=& \Biggl[\alpha^i_{-2} -{2 \over \Pplus(1)}
 \alpha^i_{-1}\alpha^+_{-1} -{p^i_f \over \Pplus(2)}\alpha^+_{-2} 
  \nn\\
 & &  + {p^i_f \over \Pplus(1)}\left\{ {1\over \Pplus(1)}
  + {1\over \Pplus(2)}\right\} (\alpha^+_{-1})^2 \Biggr]
 \pdownvac \period
\eqaend
The reader familiar with string theory recognizes that these are 
 essentially the transverse physical states of bosonic string 
 theory.  In fact this observation will allow us to find a 
  better representation of physical states in the next section. 
(Actually, in addition to these \lq\lq transverse" states, there 
 are so called \lq\lq discrete states". However, unlike in the 
$c\le 1$  models, they occur only at level 1 and will not play 
 important roles in the subsequent discussions.)
\section{ Extraction of Space-time Geometry}
\sectionnumbering
Thus far, we have been able to quantize the model exactly and 
 obtained  all the physical states characterized by the BRST-closed 
 condition $d|\Psi\rangle =0$. This condition is a quantum expression 
 of the classical constraint $T_{\mu\nu}=0$ and herein lies 
  the gravitational dynamics.  The next task is obviously to try to 
understand what physics these states describe.  But we now face a 
problem: $|\Psi\rangle$ is an  abstract state made up of free field 
oscillators and due to the general covariance it does not depend on the 
coordinates. Not a shadow of space-time picture is in sight! 
\parsmallskip
This kind of situation is in fact not uncommon in quantum mechanics.
For example, an excited state of a harmonic oscillator of the form 
$\left(a^\dagger\right)^n |0\rangle$ by itself does not tell us anything
 about its physical content. We must act on it by such physically 
 interpretable operators as the energy, the momentum, and so on and 
 see the response. Or we must compute the expectation values of such 
 operators in these states in order to form a physical picture. So we 
 must do the same for our problem; we must act on $\ket{\Psi}$ by some 
physics-probing operators in order to get out a space-time picture. 
\parsmallskip
What operators should we use ? Preferably, one would like to use 
 BRST (gauge) invariant operators and we  do have infinitely 
 many such operators at hand. These are the \lq\lq vertex operators"
 familiar in string theory. The problem is that unlike in string 
 theory, where we are interested in the physics in the target space, 
 \lq\lq vertex operators" are extremely hard to interpret in 
 the context of two-dimensional gravity. Being integrals of local 
 operators, they are coordinate independent and hence their expectation
 values are just numbers and unfortunately there is yet no scheme of 
 getting physics out of them. \parsmallskip
Another suggestion may be to use 
 gauge-invariant \lq\lq clocks" and \lq\lq rulers". While conceptually
 correct, it is virtually impossible to construct operators
 that would work like such devices out of the fields in hand. 
\parsmallskip
 As a matter of fact, even in classical general relativity, we do not 
 know how to get space-time picture by only using gauge-invariant 
 quantities. What we normally do is to first compute the 
gauge-variant quantities such as $g_{\mu\nu}, R_{\mu\nu}, R,\,  
 T^f_{\mu\nu}$ etc., and then try to extract gauge-invariant 
consequenses. What the principle of general covariance dictates is not 
that the measurable quantities are observer-independent but that 
 the behavior of quantities measured by different observers 
 is subjected to the same physical laws. \parsmallskip
For the reasons just described and in short of a better idea, we 
 shall employ the same procedure as in classical general relativity. 
 Namely, we shall choose a gauge (within the conformal gauge) and 
 compute the  mean values $\langle g_{\mu\nu} 
\rangle$ etc. in some interesting states and interpret them as what we 
see on the average in these states. In partucular, we will try to 
 construct states describing  black hole configurations in the 
 limit of large $L$ (the parameter size of the universe)
\cite{KS1}, \cite{KS2}. \parsmallskip
As we begin putting this idea into practice, we immediately face 
 a couple of problems. First stems from the fact that, 
 in the present context, the Virasoro levels express the discretized 
 energy levels of the fields, not the squared mass as in string theory. 
 If we denote the level by $n$, we have 
\eqabegin
 E_n &\propto & {n\over L} \period \label{energylevel}
\eqaend
As we wish to produce configurations where matter fields carry 
{\it finite }  energy in the limit $L \rightarrow \infty$, we need to 
controle physical states with {\it arbitrarily high Virasoro levels}. 
We will shortly describe how it can be done. \parsmallskip
Another problem is how we should define the inner product between 
 states, especially for the zero-mode sector, since it is known 
\cite{Arisetal} that hermitian operators with {\it continuous } 
spectra need not have  real eigenvalues. Thus we must be careful 
 to ensure the reality of physical observables. As this is rather 
 technical, we refer the reader to the original paper\cite{KS2} and 
 will not elaborate on it further. 
\subsection{ Physical States at Arbitrary Level}
Previously, we have obtained a compact formula (\ref{dhatcohom}) 
 for arbitrary 
 physical states in the BRST formalism. That expression, however, 
 is rather formal and will not be useful for our purposes. Now the 
 fact that these states are essentially the transverse states 
in string theory suggests a better idea; the use of DDF type spectrum 
 generating operators \cite{DDF}. Indeed we can prove that the states 
 produced by BRST formalism can be generated by 
 the following oscillators $\tilde{A}_m^i$ :
\eqabegin
 \tilde{A}_m^i &\equiv& \e^{i(m/\gt\pplus)\ln(\gt\pplus)} 
 \int_0^{2\pi}{d\yplus\over 2\pi}
 \e^{im\eta^+/(\gt\pplus)} \delplus \phi_f^i(y^+) \comma \label{Atil}
\eqaend
where $\phi^i_f$ is a normalized matter field and it is 
 \lq\lq dressed" by the field $\etap$ in the exponent. 
 It is a dimension 0 primary  defined by 
\eqabegin
 \eta^+ &=& \ln \left(\exp(\gt\qplus/2)A(\xplus)/\mu\right) \comma 
\eqaend
where $A(\xplus)$ is the operator defined before, namely 
\eqabegin
 A(\xplus) &=& \mu C(\alpha)\int_0^{2\pi} d\sigma' E_\alpha 
(\sigma-\sigma')\e^{\psi_+(\sigma')}\period
\eqaend
(The phase factor in front in (\ref{Atil}), made up of zero mode $\pplus$,
 is there to cacel a phase in the integrand.) 
It is easy to show that $\tilde{A}_m^i$'s satisfy the commutation 
 relations for oscillators, $\left[ \tilde{A}_m^i, \tilde{A}_n^j\right] 
= m\delta_{m+n,0}$ and they are BRST invariant. Thus we can build up
 physical states like $\ket{\psi} = \sum C_{m_1\ldots m_N}^{i_1 
\ldots i_N}\tilde{A}_{-m_1}^{i_1}\cdots \tilde{A}_{-m_N}^{i_N}\pdownvac
$ where $\pdownvac$ is a BRST invariant zero mode state. 
An important feature of $\tilde{A}_m^i$ is that it does not contain 
 the field $\chi$. Consequently, $\psi$ is not active upon physical 
 states built this way and it simplifies our calculations. 
\parsmallskipn
%
\subsection{ Choice of Probing Operators and Macroscopic States}
As we have already stated, we shall use gauge-variant but easily 
 interpretable operators to probe physics. Specifically, we 
will deal with 
\eqabegin
 T^f(\xiplus) &=& {1\over \gt^2 }:\left(\del_{\xiplus}
\vec{f}\right)^2 : \comma \\
g^{\alpha\beta} &=& \left(\nchiepsi + AB\e^{-\psi}\right)
\eta^{\alpha\beta} \comma 
\eqaend
where the normal ordering is defined as usual by 
\eqabegin
\nchiepsi &\equiv& \chi \e^{-\psi} -\left[\chia, \e^{-\psi}\right]
\comma 
\eqaend
with $\chia$ denoting the annihiliation part of $\chi$. 
\parsmallskip
%
Next, we must fix a gauge within the conformal gauge. Recall that 
 any physical state $\ket{\Psi}$ is of the form 
\eqabegin
 \ket{\Psi} &=& \ket{\Psizero}+d\ket{\Lambda}\comma\\
\eqaend
where $\ket{\Psizero}$ has vanishing weight and constitutes the non-trivial 
part of the cohomology. We know that a choice of $d\ket{\Lambda}$ part 
 should not change physics (as long as it will not cause anything 
 singular in various quantities). Thus it should correspond to the 
 gauge freedom left in the conformal gauge. Let us give an example. 
Let ${\cal O}$ be a $(1,1)$ operator like the metric $g_{\mu\nu}$. Under
 the  conformal transformation, it transforms as $\delta g_{\mu\nu} 
= \nabla_\mu \epsilon_\nu +\nabla_\nu \epsilon_\mu$. 
For $\xplus \longrightarrow \xplus +\epsilon^+(\xplus)$ in conformal
 gauge, this becomes 
\eqabegin
 \delta g_{+-} &=& \nabla_+\epsilon_- =
\delplus\epsilon_- \comma \label{gtofg}
\eqaend
 since only non-vanishing component of 
 the Christoffel symbol is   $\Gamma^+_{++}$. 
Now under $\ket{\Lambda} \rightarrow \ket{\Lambda}+
\ket{\delta\Lambda}$, the expectation value of ${\cal O}$ changes 
 by 
\eqabegin
\delta <{\cal O}(x) > &=& \bra{\Psi}\calO (x) d\ket{\delta\Lambda}
+h.c.\nn\\
&=& -\bra{\Psi}\left[d,\calO (x) \right]\ket{\delta\Lambda}
+h.c.\nn\\
&=&-\bra{\Psi}\sum_n c_{-n}\left[L_n ,\calO (x) \right]\ket{\delta\Lambda}
+h.c.\nn\\
&=& \delplus\left( -\bra{\Psi}c(\xplus)\calO(x)\ket{\delta\Lambda}
+h.c.\right) \comma 
\eqaend
 and this is precisely of the form (\ref{gtofg}) of a gauge 
transformation. \parsmallskip
What would be a convenient  choice for the 
gauge part $\ket{\Lambda}$?   
A hint is provided by the following simple observation. Let 
 $\ket{\psi_n}$ be a state with Virasoro weight $n$, \ie 
 $L_0^{tot}\ket{\psi_n}=n\ket{\psi_n}$. Then for any operator 
${\cal O}(\xplus)$ carrying a global weight, we have 
\eqabegin
 \bra{\psi_n}\left[L_0^{tot}, \calO (\xplus)\right]\ket{\psi_m} 
 &=& \overi \delplus \bra{\psi_n}\calO (\xplus)\ket{\psi_m} \nn\\
 &=& (n-m)\bra{\psi_n}\calO (\xplus)\ket{\psi_m}\period
\eqaend
This gives 
\eqabegin
 \bra{\psi_n}\calO (\xplus)\ket{\psi_m}&=& c_{nm}\e^{i(n-m)\xplus}
\comma 
\eqaend
where $c_{nm}$ is a constant. This implies that (i) $\bra{\Psizero}
 \calO(\xplus)\ket{\Psizero}$ can only be a constant and (ii) if 
 we wish to produce various coordinate dependence, we should take 
 the gauge part to be a superposition of states with various
 Virasoro levels. \parsmallskip
With this in mind, we shall take our physical state to be of the 
 form (now including both left and right moving modes) 
\eqabegin
 \ket{\Psi} &=& \ket{\Psi_0} + {1\over \kappa}\left( d\, b_{-M}
 + \bar{d}\, \bar{b}_{-M}\right)\ket{\Omega} \comma 
\eqaend
where $b_{-M}\ (\bar{b}_{-M})$ is the left (right) going anti-ghost
 at level $M$ and the constant $\kappa$ will be taken to be of order
 $ \hbar^2$. The state $\ket{\Omega}$, which constitutes the 
 gauge part, is made up only of zero-modes (zero-modes do carry 
 Virasoro weights), and is give by 
\eqabegin
\Omegaket & \equiv & \sum_{k= - \infty }^{\infty} \omega_k 
\sum_{l= \pm 1 ,0}
           \mid \tilde{P}(k,l)>   \comma \\
\mid \tilde{P}(k,l)>  & \equiv & \, e^{-ic p^+/\hbar \gamma}
  \gamma^2 \int_{-\infty}^{\infty} dp^+ 
  \int_{- \infty}^{\infty} dp^- \, \weight \nn\\
 & & \cdot \delta (p^- - \frac1{p^+}(\hbar - \half p_f^2)) 
\mid p^+,p^-(k,l),\vec{p}_f  >  
   \comma \\
 p^- (k,l) & \equiv & p^- - \frac{\hbar}{p^+} k  - 
       i \hbar \gt\, l \period
\eqaend
It may look a bit complicated but it satisfies minimum requirements 
 for a good gauge part. Let us briefly explain various factors:
$\omega_k$ is a constant, the exponential factor in front of the 
 integral is BRST invariant and will be used to cancel certain 
unwanted terms in the mean value $\langle g^{\mu\nu} \rangle$, 
 $W(\pplus)$ is a smearing function to make certain integrals finite, 
 the $\delta$-function enforces the zero-energy condition for the 
 zero mode state with $(k,l)=(0,0)$, and the two types of shifts from 
$\pminus$ in the definition of $\pminus(k,l)$ are needed to make certain 
 mean values non-trivial and real. $\omega_k$ and $W(\pplus)$ will be 
 given explicitly later. \parsmallskip

Now as for the non-trivial part $\ket{\Psizero}$, we want it to be
 able to describe states where matter shock wave with finite energy 
 is present and produces a black hole. Such a macroscopic configuration
 should be naturally described by a coherent state. Thus, by using 
 the physical oscillators $\tilde{A}_n$ (for simplicity, we consider 
 cases where only one of the matter fields is excited and hence drop the 
superscript), we define 
\eqabegin
\Psizeroket & \equiv & e^G \psmearvac\comma    \qquad 
G \equiv  {1\over \hbar}\sum_{n\ge 1}{ \tilde{\nu}_n\over n}
\Atil_{-n}  \comma \\
\tilde{\nu}_n &=& \nu_n \e^{in x^+_0 }
\qquad (\nu_n,\, x^+_0\,:\, \mbox{real constants})  \comma \\
\psmearvac &\equiv& \mid \tilde{P}(0,0)>\period
\eqaend
By construction $\Psizeroket$ is indeed a coherent state and 
satisfies $\Atil_n \Psizeroket = \tilde{\nu}_n\Psizeroket$. 
$x^+_0$ will turn out to specify the location of the shock wave. 
%
\subsection{Emergence of Black Hole Geometry}
Having prepared the physical states, we can start computing 
 the average values of the physics-probing operators of our 
 choice $g^{\mu\nu}, T^f(\xplus)$ in such states. 
Computations are long but can be carried out exactly. 
The results, with the parameters such as $\omega_k, \nu_n$ etc. 
 left unspecified, still contain infinite sums and look 
 more or less like a mess. We shall not display them. 
However, as we choose our parameters appropriately and take the 
 large $L$ limit, drastic simplification occurs and one begins to 
 see a picture of the averaged space-time. The reason for this 
 simplification is that the terms expressing contributions 
 from modes at finite Virasoro levels all die away in this limit and 
 furthermore the infinite sums can be replaced by tractable integrals. 
\parsmallskip
Let us now describe what we get. First we must specify the 
 parameters concretely. Since the energy variable 
$u\equiv n/L$ tends to be continuous as $L\rightarrow \infty$, 
 dependence on the level $n$ should be replaced by that on $u$. 
With this in mind we choose the parameters as 
\eqabegin
 \nu_n &=& \nu(Lu) = \nu u^d \e^{-au^2} \comma \\
 \omega_n &=& \omega(Lu) = {-\omega \over Lu} \qquad (n \ne 0) 
\comma \\
 \omega &=& \mbox{ a positive constant}  \comma \\
 \omega_0 &=& \mbox{a constant to be adjusted } \comma \\
 \weight &=& \pplus \e^{-\alpha(\pplus-\pplus_0)^2/2} \period
\eqaend
In the expression for $\nu_n$, $d$ and $a$ are constants and they 
controle, respectively, the type of matter distribution and the 
 width of its flux. \parsmallskip
The most typical configuration turned out to be produced by the 
 choice $d=-1/2$. In this case, the average behavior of the 
 left-going matter energy-momentum tensor and that of the inverse
 metric are give by 
\eqabegin
<\,T^f (\xi^+) \,>&\stackrel{L \rightarrow \infty}{\sim}&
    \omega \nu^2 I_T(\xiplus -\xiplus_0 )
  \, \psmearnorm     \comma \\
<\, g^{-1} \, > &\equiv & <\, : \left( \chi+AB\right)\e^{-\psi} :\,> 
 \nn\\ 
 &\stackrel{L \rightarrow \infty}{\sim}&
 \left(\kf(\xiplus)+\lm^2 \xiplus\ximinus \right) \psmearnorm \period
\eqaend
In the expressions above, $\lm^2$ is a constant 
 proportional to $\lambda^2$,  the common factor 
$\psmearnorm$ is the norm of the 
zero-mode state, which can be made finite, 
the function $\kf(\xiplus)$ is given by 
\eqabegin
 \kf(\xiplus) &=& c_K\xiplus + d_K -\gt^2\omega\nu^2 I_\chi(\xiplus
 -\xiplus_0) \qquad (c_K, d_K\ \mbox{ :constants}) \comma 
\eqaend
and $I_T(\xi)$ and $I_\chi(\xi)$ are integrals of the form 
\eqabegin
 I_\chi(\xi) &=& \int_{1/L}^\infty du {\cos u\xi \over u^2}
\int_0^u dv \left[v(u-v)\right]^{-1/2} \e^{-a(v^2 + (u-v)^2)} 
 \comma \\
I_T(\xi) &=&  \int_{1/L}^\infty du \cos u\xi 
\int_0^u dv \left[v(u-v)\right]^{-1/2} \e^{-a(v^2 + (u-v)^2)} 
 \period
\eqaend
It is easy to see that these integrals are related by 
\eqabegin
 I_T(\xi) &=& -\del_\xi^2 I_\chi(\xi) \period
\eqaend
This actually expresses the fundamental energy balance condition and 
 is a good check on our calculations. To see this, recall the 
 energy-momentum constraint:
\eqabegin
\gt^2 T_{++} &=& \delplus\chi\delplus\psi -\del^2_+\chi +\half 
 (\delplus f)^2 =0\period
\eqaend
As we remarked before, on the physical states made up of $\tilde{A}_n$
 the field $\psi$ is inactive and can be regarded to vanish. Thus 
 the constraint takes the form
\eqabegin
 \half (\delplus f)^2 &=& \del^2_+\chi\comma 
\eqaend
whichi is nothing but the relation between the integrals above. 
\parsmallskip
As for the evaluation of these integrals, one can perform the 
 oscillatory integral over $u$ analytically and then do the remaining 
 integral by numerical methods. The results are plotted in Fig.8a 
 and 8b.\parbigskipn 
\qquad \qquad 
\begin{minipage}{2.5in}
\epsfxsize=5cm
\epsfysize=5cm
\epsfbox{sorf8a.ai}
\end{minipage}
\qquad\qquad 
\begin{minipage}{2.5in}
\epsfxsize=5cm
\epsfysize=5cm
\epsfbox{sorf8b.ai}
\end{minipage}\parbigskipn
\qquad \qquad 
\begin{minipage}{2.5in}
{\small{\bf Fig.8a}\quad Plot of the integral $I_T(\xi)$.}
\end{minipage}
 \quad 
\begin{minipage}{2.5in}
{\small{\bf Fig.8b}\quad Plot of the integral $I_\chi(\xi)$.}
\end{minipage}
\parbigskipn
%
$I_T(\xi)$ is very close to a Gaussian,  width of which is controled 
 by the parameter $a$ and for small $a$ it approaches a 
$\delta$ function and describes a matter shock wave. As for 
$I_\chi(\xi)$ (up to a constant $L\pi$, which can be 
 canceled by a constant $d_K$), it behaves roughly like $-(\pi^2)
 |\xi|$ except near the origin and as $a\rightarrow 0$ it converges to
  that function. Remarkably, these functions behave exactly like the 
 corresponding ones in the classical CGHS model in the limit of 
 small $a$ ! \parsmallskip
One then expects that a black hole is generated due to the matter 
 flux. To confirm this, let us compute the curvature scalar. 
 It is given by 
\eqabegin
 R^g 
 &=&4\lm^2 {\kf(\xiplus) -\xiplus \delplus \kf(\xiplus)
 \over \kf(\xiplus) + \lm^2 \xiplus\ximinus } 
\period
\eqaend
With appropriate choices of parameters, $\kf$ takes the form
\eqabegin
 \kf(\xiplus) &=& \gt^2\omega \nu^2 \left( {\pi^2\over 2}
 (\xiplus -\xiplus_0) -I_\chi(\xiplus -\xiplus_0)+L\pi \right) 
\comma  
\eqaend
which for $a<< (\xi^+ -\xi^+_0)^2$ goes like 
\eqabegin
\kf(\xiplus) &\sim &  (\pi^2/2)\gt^2\omega \nu^2 (\xiplus -\xiplus_0)
\theta(\xiplus -\xiplus_0) \period
\eqaend
The line of curvature singularity resulting from this expression is 
 plotted in Fig.9.
\begin{center} 
\epsfxsize=8cm
\quad\epsfbox{sorf9.ai}
\end{center}
\begin{center}
{\small {\bf Fig.9} \quad Curvature singularity for $d=-1/2$.}
\end{center}
\parbigskipn
It clearly describes the formation of a  black hole by a matter shock 
wave just as in the CGHS model.  \parsmallskip
If we change the value of the parameter $d$ controling the distribution
 of the matter field, we can get different configurations. For example, 
 if we take $d=1/2$, it turned out that we get a space-time 
 with a naked singularity depicted in Fig.10. 
\begin{center} 
\epsfxsize=8cm
\quad\epsfbox{sorf10.ai}
\end{center}
\begin{center}
{\small {\bf Fig.10} \quad Curvature singularity for $d=1/2$.}
\end{center}
\parbigskip
We would like to emphasize that although the space-time pictures
 that emerged in the examples above look \lq\lq classical" actually all the 
 quantum corrections have been fully taken into account.  In fact 
 the very notion of \lq\lq quantum correction" becomes ambiguous in 
 exact treatment since we do not have a classical background to begin 
 with. As is apparent, the coherent state $\ket{\Psizero}$ is a highly
 non-perturbative quantum state and precisely through its {\it quantum 
 coherence} macroscopic classical-like configurations are formed. 
\section{ Future Problems}
\sectionnumbering
We hope to have convinced the reader that   
our attempt, despite the use of a simple model with a number of 
shortcomings, has produced certain success. Namely, we have shown
 that a certain model of CGHS type can be quantized exactly and 
 further  we have explicitly implimented an idea of getting 
 physical pictures  out of abstract physical states. On the other hand,
 our analysis  also illustrated the nature of 
 difficulties that one must face in any challange for exact treatment of 
quantum gravity. We now list some of the important  problems left
 unsolved. 
\parsmallskip
The most serious among them is {\it how to define and compute the 
$S$-matrix }, which is a prerequisite for proper analysis of the  
Hawking radiation.  First, the definition of scattering matrix calls 
 for a separation of bulk geometry and particle excitations. In the 
 semi-classical treatment, this is not a serious problem since the 
 space-time picture (\ie the background geometry) is already available 
 {\it before} one starts discussing the $S$-matrix. On the other hand, 
 in an exact treatment, \lq\lq geometry" emerges only 
{\it after} the computation of some expectation values. This 
 is a universal difficulty and not singular to our particular treatment 
since by the very requirement of general covariance the physical states 
 by themselves cannot contain any dependence on the coordinates. 
\parsmallskip
Deeply related to the point above is the fact that the definition of 
 the  $S$-matrix requires a separation of sub-systems or sub-regions. 
 For instance, the $S$-matrix is supposed to relate the states in the 
\lq\lq far past" to those in the \lq\lq far future" but the notion 
 of \lq\lq time" itself is, to say the least, hard to come by. 
\parsmallskip
Another famous problem one must face in an attempt for  exact 
quantization 
 is the question of quantum (in)coherence. Logically, pure states 
 cannot possibly evolve into mixed states, as is the case in our 
 treatment. This would mean that the apparent quantum incoherence 
 implied by the thermal spectrum of Hawking radiation can only be 
 due to the semi-classical approximation used in deriving it. 
 As this is a general problem in the quantum mechanics of 
 macroscopic objects in interaction with microscopic states, it 
 would be interesting to construct a simple model to demonstrate this. 
\parsmallskip
Finally, let us mention the problem of how to define, at least conceptually, 
 the notion of \lq\lq quantum states of a black hole". This 
 is again a very hard problem in the exact treatment since at present
 one has no idea how to characterize a quantum state which contains 
 a black hole. One would have to somehow bring in the notion of 
causal structure at the abstract level. Any such idea would be 
 extremely interesting since, as is well known, this problem is 
 directly related to the question of the statistical meaning of 
 the black hole entropy and of the no hair theorem. 
\parsmallskip
We hope that some of these problems can be solved, at least in 
 a clever model, in the near future. \parbigskipn\parbigskipn
 {\Large\bf Acknowledgment } \parbigskip
I would like to thank the organizers of the symposium, especially 
 J.E. Kim and C.K. Lee,  for their hospitality and for providing an
 excellent atmosphere for the academic  (and additional)  activities.
This work is supported  in part by the Grant-in-Aid for Scientific 
Research (No. 06640378) and  Grant-in-Aid for Scientific Research for 
Priority Area (No. 06221211) from the Ministry 
 of Education, Science and Culture.
\parbigskipn\parbigskipn

\end{document}